\documentclass[aps,prd,onecolumn,superscriptaddress,longbibliography]{revtex4-2}
\usepackage{graphicx}
\usepackage{epstopdf}
\usepackage{epsfig}
\usepackage{amsmath}
\usepackage{color}
\usepackage{amssymb}  
\usepackage{amsfonts} 
\usepackage{amsbsy}   
\usepackage{slashed}
\usepackage{bm}
\usepackage{hyperref}
\hypersetup{
    colorlinks=true,       
    linkcolor=red,        
    citecolor=blue,        
    urlcolor=blue          
}

\newcommand{\beq}{\begin{align}}
\newcommand{\eeq}{\end{align}}
\newcommand{\ba}{\begin{array}}
\newcommand{\ea}{\end{array}}
\newcommand{\bea}{\begin{eqnarray}}
\newcommand{\eea}{\end{eqnarray}}
\newcommand{\bi}{\begin{itemize}}
\newcommand{\ei}{\end{itemize}}
\newcommand{\ben}{\begin{enumerate}}
\newcommand{\een}{\end{enumerate}}
\newcommand{\bc}{\begin{center}}
\newcommand{\ec}{\end{center}}
\newcommand{\bl}{\begin{flushleft}}
\newcommand{\el}{\end{flushleft}}
\newcommand{\br}{\begin{flushright}}
\newcommand{\er}{\end{flushright}}

\newcommand\comment[1]{ \hbox{[{\it Comment suppressed here.}\/]} }
\newcommand\hide[1]{}
\newcommand{\skipover}[1]{}

\begin{document}

\title{Mass spectra of charged mesons and the quenching of vector meson condensation via exact phase-space diagonalization}

\author{Jingyi Chao}%
\email{chaojingyi@jxnu.edu.cn}
\affiliation{ 
College of Physics, Jiangxi Normal University, Nanchang, Jiangxi 330022, China}
\author{Kun Xu}%
\email{xukun@bit.edu.cn}
\affiliation{School of Physics, Beijing Institute of Technology, Beijing 100080, China}

\date{\today}

\begin{abstract}
We investigate the dynamics and mass spectra of charged pseudoscalar ($\pi^+$) and vector ($\rho^+$) mesons in a background magnetic field at finite temperature using the two-flavor Nambu-Jona--Lasinio (NJL) model. By employing a quark propagator that isolates the Schwinger phase from its Landau level expansion, we formulate an exact non-commutative phase-space framework utilizing the Wigner-Weyl transform and the Moyal star product. This approach enables the algebraic diagonalization of the Bethe-Salpeter equations for composite states with asymmetric fractional constituent charges. For the pseudoscalar channel, we analytically verify the exact cancellation between the dynamical random phase approximation spatial sum rules and the vacuum gap equation. This identity preserves the generalized Goldstone theorem, causing the $\pi^+$ pole mass to strictly track the kinematic zero-point energy drift at order of $eB$. In the vector channel, our full phase-space evaluation reveals that the Zeeman spin-splitting emerges dynamically from microscopic threshold truncations governed by the chiral Dirac algebra. Notably, we find that the tachyonic instability of the spin-aligned $\rho^+$ state is quenched. The magnetic catalysis of the chiral condensate drives the continuum threshold ($2M$) upwards, overtaking the Zeeman attraction and preventing vector meson condensation within this mean-field framework. Furthermore, finite-temperature evaluations show a monotonic thermal suppression of the meson masses driven by Pauli blocking, yet all modes remain bound without undergoing Mott dissociation prior to chiral symmetry restoration.
\end{abstract}

\maketitle

\section{Introduction}
\label{sec:intro}

Understanding the phase structure and dynamic properties of strongly interacting matter under extreme conditions is a central endeavor in modern quantum chromodynamics (QCD)~\cite{Kaspi:2017fwg}. In recent years, the physics of QCD matter exposed to intense background magnetic fields has attracted significant theoretical and experimental interest \cite{Kharzeev:2012ph, Miransky:2015ava, Andersen:2014xxa}. Such extreme magnetic environments are physically realized in various systems, ranging from the dense cores of magnetars ($eB \sim 10^{-4}-10^{-2} \text{ GeV}^2$) to relativistic heavy-ion collisions at the Relativistic Heavy Ion Collider (RHIC) and the Large Hadron Collider (LHC), where transient magnetic fields can reach extraordinary magnitudes on the hadronic scale ($eB \sim 0.1-1.0 \text{ GeV}^2$) \cite{Skokov:2009qp, Deng:2012pc}. The presence of these intense magnetic fields explicitly breaks spatial rotational symmetry and modifies the vacuum structure, leading to non-perturbative phenomena such as the Chiral Magnetic Effect (CME) \cite{Fukushima:2008xe} and Magnetic Catalysis (MC) of chiral symmetry breaking \cite{Gusynin:1995nb, Bali:2012zg, Bruckmann:2013oba, Chao:2013qpa, Shovkovy:2012zn, Ding:2026qzu}.

Among these magnetic phenomena, the mass spectrum and structural evolution of charged composite hadrons, specifically the pseudoscalar pion ($\pi^\pm$) and the vector rho meson ($\rho^\pm$), serve as fundamental probes of the magnetized QCD vacuum. For the $\pi^\pm$ meson, its mass behaves as $m_{\pi^\pm}^2 \approx m_\pi^2 + |eB|$ in the weak-to-moderate field regime \cite{Agasian:2001ym, Andersen:2012dz, Orlovsky:2013gha, Colucci:2013zoa, Avancini:2016fgq, Brauner:2016pko, Bali:2018sey, Brandt:2018bwq, Ding:2020hxw, Xing:2021kbw, Wen:2024hgu, Wang:2026xsm}. However, the dynamics of the charged vector meson $\rho^+$ present a more controversial theoretical landscape, its energy level is governed by $m^2_{\rho^{\pm}} = m_\rho^2 + (2n + 1 - 2S_z)|eB|$ in the point-particle level \cite{Hidaka:2012mz, Kawaguchi:2015gpt, Andreichikov:2016ayj}. The spin-aligned ground state ($n=0, S_z=1$) predicts a severe Zeeman attraction $m_{\rho^+}^2 \approx m_\rho^2 - |eB|$. Consequently, at a critical magnetic field $eB_c \approx m_\rho^2$, the macroscopic mass is predicted to plunge into the tachyonic regime, triggering Vector Meson Condensation (VMC) \cite{Callebaut:2010mct, Chernodub:2010qx, Chernodub:2011mc, Brauner:2016lkh}.

In reality, the charged $\rho$ meson is a composite quark-antiquark ($u\bar{d}$) bound state governed by QCD dynamics at low energy scale. Recent first-principle Lattice QCD (LQCD) simulations \cite{DElia:2011koc, Luschevskaya:2014lga, Bali:2017ian, DElia:2018xwo, Ding:2020jui, Endrodi:2024cqn} and effective field theory evaluations \cite{Wang:2017vtn} indicate that while the $\rho^\pm$ mass initially decreases, it ultimately avoids the tachyonic zero. These advancements emphasize that internal quark structure and interactions with the QCD vacuum, such as magnetic catalysis, are critical for stabilizing the vector meson. To describe these composite dynamics, models such as the Nambu-Jona--Lasinio (NJL) model \cite{Klevansky:1992qe, Liu:2018zag, Chaudhuri:2019lbw, Avancini:2021pmi, Cao:2021rwx, Sheng:2020hge, Xu:2020yag, Chao:2022bbv, Yuan:2023dco} are widely used. However, when evaluating the full double summation or integration over the constituent quarks' internal momenta, traditional approximations, such as the lowest Landau level or symmetric ultraviolet cutoffs, encounter certain challenges. Specifically, they may not fully preserve exact spatial sum rules, potentially introducing theoretical artifacts into the Bethe-Salpeter (BS) equations~\cite{Chang:2009zb}. Alternatively, the Ritus eigenfunction method \cite{Ritus:1972ky, Zhang:2016qrl} is a standard framework for evaluating fermionic observables, with ongoing efforts exploring its formal equivalence to the Schwinger scheme \cite{ConcurrentWork2026}. On the other hand, due to the asymmetric fractional charges of the constituent $u$ and $\bar{d}$ quarks, the Ritus scheme requires coupled eigenfunctions across both coordinate and momentum spaces, resulting in complicated expressions.

Fundamentally, to avoid those analytical complexities and address the charge geometry more directly, we factorize the gauge-dependent Schwinger phases and map the tensor Bethe-Salpeter equations into Wigner-Weyl phase space \cite{Moyal:1949sk}. Within this non-commutative framework, the Moyal star product enables an algebraic diagonalization of the scattering matrix across the asymmetric $(n_u, n_d)$ constituent quark Landau levels \cite{Moyal:1949sk, Curtright:1997me, Zachos:1999wn}. This projection reduces the spatial convolutions to generalized triple Laguerre overlap integrals, explicitly conserving probability and transverse kinetic energy. Moreover, because the spatial treatment cleanly decouples from the Lorentz spin structure, the method applies uniformly to both pseudoscalar and vector charged mesons. We apply this exact phase-space diagonalization to re-evaluate the mass spectra of the magnetized $\rho^\pm$ triplet states, revealing microscopically how chiral dynamics and threshold truncations inherently quench the tachyonic instability.

The paper is organized as follows. In Sec.~\ref{sec:ch1_gap_ds}, we introduce the magnetized NJL model, formulate the exact gauge-phase factorization, and derive the exact vacuum gap equation. In Sec.~\ref{sec:ch2_pion}, we apply the Wigner-Weyl projection and the Moyal star product to the pseudoscalar channel, analytically demonstrating the Goldstone protection mechanism and numerically evaluating the $\pi^+$ mass. In Sec.~\ref{sec:ch3_rho}, we extend the tensor non-commutative framework to the vector channel, exploring the dynamically generated Zeeman splitting, the quenched VMC mechanism, and the finite-temperature mass evolution. Finally, we summarize our findings and conclude in Sec.~\ref{sec:summary}. Technical details regarding the analytical proof of the projector algebra and the exact Laguerre overlap sum rules are relegated to the Appendices.

\section{Magnetized Quark Propagators and Phase-Space Quantization}
\label{sec:ch1_gap_ds}

We use the two-flavor NJL model to investigate the dynamics of charged mesons, the Lagrangian of which in the presence of  background magnetic field is given by \cite{Klevansky:1992qe}:
\begin{equation}
    \mathcal{L} = \bar{\psi} (i \slashed{D} - \hat{m}) \psi + G \left[ (\bar{\psi} \psi)^2 + (\bar{\psi} i \gamma_5 \boldsymbol{\tau} \psi)^2 \right] - G_V \left[ (\bar{\psi} \gamma^\mu \boldsymbol{\tau} \psi)^2 + (\bar{\psi} \gamma^\mu \gamma_5 \boldsymbol{\tau} \psi)^2 \right],
\end{equation}
where $\psi = (u, d)^T$ represents the quark field doublet with $\hat{m} = \mathrm{diag}(m_u, m_d)$ the current quark mass, and $\boldsymbol{\tau}$ are the Pauli matrices in flavor space. The gauge-covariant derivative is defined as $D^\mu = \partial^\mu + i \hat{Q} A^\mu$, which couples the quarks to the external electromagnetic vector potential $A^\mu$ via the fractional charge matrix $\hat{Q} = \mathrm{diag}(q_u, q_d) = \mathrm{diag}(\frac{2}{3}e, -\frac{1}{3}e)$. The constants $G$ and $G_V$ denote the effective four-fermion coupling strengths in the scalar-pseudoscalar and vector-axial-vector channels, responsible for generating the $\pi$ and $\rho$ meson bound states, respectively.

In a constant homogeneous magnetic field $\boldsymbol{B} = B \hat{z}$, full translational symmetry is explicitly broken. The coordinate-space quark propagator can be factorized into a gauge-dependent Schwinger phase and a translationally invariant, gauge-independent part \cite{Schwinger:1951nm, Chao:2014wla, Miransky:2015ava, Sheng:2017lfu}:
\begin{equation}
    S_f(x, x') = e^{i\Phi_f(x, x')} \int \frac{d^4p}{(2\pi)^4} e^{-i p \cdot (x - x')} \tilde{S}_f(p),
\end{equation}
where $\Phi_f(x, x') = \frac{1}{2} q_f x_\mu F^{\mu\nu} x'_\nu$ is the Schwinger phase in the symmetric gauge, with $F^{\mu\nu}$ being the electromagnetic tensor. In this exact coincidence limit, the gauge-dependent Schwinger phase $\Phi_f(x,x)$ strictly vanishes, preserving macroscopic gauge invariance. The translationally invariant momentum-space propagator $\tilde{S}_f(p)$ can be expressed as an infinite sum over internal fermionic Landau levels:
\begin{equation}
    \tilde{S}_f(p) = i e^{-\alpha_f} \sum_{n=0}^\infty \frac{\slashed{D}_n(q_f B, p)}{p_\parallel^2 - M^2 - 2n|q_f B| + i\epsilon},
\end{equation}
where $\alpha_f \equiv \boldsymbol{p}_\perp^2 / |q_f B|$. The numerator $\slashed{D}_n(q_f B, p)$ enclose the spin dynamics and the dimensionally reduced Dirac structures within each Landau level:
\begin{equation}
    \slashed{D}_n(q_f B, p) = (\gamma_\parallel \cdot p_\parallel + M) \left[ \mathcal{P}_+^f (-1)^n L_n(2\alpha_f) + \mathcal{P}_-^f (-1)^{n-1} L_{n-1}(2\alpha_f) \right] + 2 (\boldsymbol{\gamma}_\perp \cdot \boldsymbol{p}_\perp) (-1)^{n-1} L_{n-1}^1(2\alpha_f).
\end{equation}
Here, $p_\parallel = (p_0, 0, 0, p_z)$, $\boldsymbol{p}_\perp = (0, p_x, p_y, 0)$, and the spin projection operators are defined as $\mathcal{P}_\pm^f = \frac{1}{2}(1 \pm i s_f \gamma^1 \gamma^2)$ with $s_f = \mathrm{sgn}(q_f B)$. For notational brevity, we define the uncharged projectors as $\mathcal{P}_\pm = \frac{1}{2}(1 \pm i \gamma^1 \gamma^2)$. Note that the generalized Laguerre polynomials $L_{-1}^{(a)} \equiv 0$, which effectively annihilates the unphysical spin states in the lowest Landau level ($n=0$).

Under mean-field approximation, the quark  constituent quark mass $M$ is determined by the self-consistent gap equation \cite{Klevansky:1992qe, Andersen:2014xxa}:
\begin{equation}
    M = m_0 - 2G \langle \bar{\psi}\psi \rangle = m_0 - 2G \sum_{f=u,d} \langle \bar{f}f \rangle,
\end{equation}
where $m_0$ is the current quark mass (assuming isospin symmetry $m_u = m_d = m_0$), and the condensate:
\begin{equation}
    \langle \bar{f}f \rangle = -i N_c \mathrm{Tr}_{\mathrm{D}} \left[ \lim_{x' \to x} S_f(x, x') \right],
\end{equation}
where $\mathrm{Tr}_{\mathrm{D}}$ denotes the trace over the Dirac spinor indices, and the explicit factor $N_c$ arises from the trace over the color space. Substituting the Landau level expansion and evaluating the Dirac trace yields the fundamental vacuum structure. We identify the universal chiral scalar density $J_1$, which determines the gap equation. Summing over the flavors, it reads:
\begin{equation}
    J_1 = \sum_{f=u,d} \frac{1}{2} \left[ -i N_c \frac{|q_f B|}{2\pi} \sum_{n_f=0}^\infty \beta_{n_f} \int \frac{dp_0 dp_z}{(2\pi)^2} \frac{1}{p_0^2 - p_z^2 - M^2 - 2n_f|q_f B| + i\epsilon} \right],
\end{equation}
where $\beta_{n_f} = 2 - \delta_{n_f, 0}$ explicitly accounts for the spin degeneracy of the Landau levels. This dimensionally reduced integral directly relates to the chiral condensate via $J_1 = \sum_{f} \langle \bar{f}f \rangle_B / (8M)$.

When constructing the coordinate-space polarization tensor for the charged pion $\pi^+$ , we trace over the closed quark loop:
\begin{equation}
    \Pi_{\pi^+}(x, y) = -i N_c \mathrm{Tr}_{\mathrm{D}}\left[ \gamma_5 S_u(x, y) \gamma_5 S_d(y, x) \right],
\end{equation}
Since $\Phi_d(y, x) = -\Phi_d(x, y)$, the total accumulated Schwinger phase in the loop evaluates to:
\begin{equation}
    \Phi_{\mathrm{loop}}(x, y) = \Phi_u(x, y) + \Phi_d(y, x) = \frac{1}{2} (q_u - q_d) x_\mu F^{\mu\nu} y_\nu = \frac{1}{2} q_\pi x_\mu F^{\mu\nu} y_\nu \equiv \Phi_\pi(x, y).
\end{equation}
Given the fractional charges $q_u = \frac{2}{3}e$ and $q_d = -\frac{1}{3}e$, this residual phase elegantly matches the macroscopic Schwinger phase of a composite particle carrying the pion's physical net charge $q_\pi = e$. 

To determine the mesonic pole, we consider the BS integral equation for the scattering $T$-matrix in the Random Phase Approximation (RPA) channel \cite{Klevansky:1992qe, Avancini:2015ady, Mao:2018dqe}:
\begin{equation}
    T(x, y) = 2G \delta^4(x-y) + \int d^4z \, K(x, z) T(z, y),
\end{equation}
where the first term is the local bare contact interaction, and $K(x, z) \equiv 2G \Pi(x, z)$ represents the irreducible one-loop polarization kernel. Because both $T$ and $K$ describe the propagation of the charged pion, they share the identical global Schwinger phase: $T(x, y) = e^{i\Phi_\pi(x, y)} \tilde{T}(x-y)$ and $K(x, y) = e^{i\Phi_\pi(x, y)} \tilde{K}(x-y)$. Particularly, the local bare vertex inherently satisfies this factorization since $2G \delta^4(x-y) = e^{i\Phi_\pi(x, y)} [2G \delta^4(x-y)]$ trivially holds at $x=y$. When performing the spatial convolution over the intermediate coordinate $z$, the product of the gauge phases satisfies the geometric identity~\cite{Schwinger:1951nm}:
\begin{equation}
    \Phi_\pi(x, z) + \Phi_\pi(z, y) = \Phi_\pi(x, y) + \frac{q_\pi}{2} (x-z)_\mu F^{\mu\nu} (z-y)_\nu.
\end{equation}
By factoring out the global phase $e^{i\Phi_\pi(x, y)}$ from the entire equation, this relation explicitly demonstrates that the coordinate difference inside the integral introduces the additional phase factor $e^{i \frac{q_\pi}{2} (x-z)_\mu F^{\mu\nu} (z-y)_\nu}$.

Upon performing a Fourier transform to momentum space, this phase factor translates into momentum derivatives. The standard convolution integral thereby morphs into the non-commutative Moyal star product \cite{Moyal:1949sk}:
\begin{equation}
    \tilde{T}(q) = 2G + \tilde{K}(q) \star \tilde{T}(q),
\end{equation}
where the bare contact interaction $2G$ drives the lowest-order scattering, and $\tilde{K}(q) \equiv 2G \tilde{\Pi}(q)$ denotes the translationally invariant part of this kernel in momentum space. The star product is formally defined through the infinite-order pseudo-differential operator:
\begin{align}
    f(q) \star g(q) &= f(q) \exp\left( -\frac{i}{2} \overleftarrow{\partial}_{q_\mu} (q_\pi F^{\mu\nu}) \overrightarrow{\partial}_{q_\nu} \right) g(q) \nonumber \\
    &= f(q)g(q) - \frac{i}{2} q_\pi F^{\mu\nu} \left[ \partial_{q_\mu} f(q) \right] \left[ \partial_{q_\nu} g(q) \right] + \mathcal{O}(F^2).
\end{align}

By expanding the exponential operator, the physical transition from the vacuum to the magnetized medium becomes transparent. The zeroth-order term trivially recovers the ordinary commutative product, strictly reproducing the standard RPA equation in the $B$-field free limit. However, the exact retention of the Schwinger phase introduces an infinite tower of higher-order momentum derivatives driven by the background field $F^{\mu\nu}$. These derivative terms entangle the transverse momenta, preventing the standard algebraic factorization of the scattering matrix. To rigorously diagonalize this non-commutative system without compromising the spatial geometry, we must project the equation onto an exact eigenbasis of the star product. 

To achieve this, we expand the kernels in terms of the basis functions $\mathcal{P}_n(\alpha_\pi) \equiv 2 e^{-\alpha_\pi} (-1)^n L_n(2\alpha_\pi)$, where $\alpha_\pi \equiv \boldsymbol{q}_\perp^2 / |q_\pi B|$. These functions act as algebraic projection operators under the Moyal star product, satisfying the non-commutative projector algebra~\cite{Berra-Montiel:2024ubb}:
\begin{equation}
    \mathcal{P}_n(\alpha_\pi) \star \mathcal{P}_{n'}(\alpha_\pi) = \delta_{n, n'} \mathcal{P}_n(\alpha_\pi).
\end{equation}
This orthogonality can be intuitively understood via the Weyl-Wigner correspondence. The basis functions $\mathcal{P}_n(\alpha_\pi)$ are the exact Wigner transforms of the one-dimensional quantum harmonic oscillator density matrices $\hat{\rho}_n = |n\rangle \langle n|$. Since these Hilbert space projection operators are strictly orthonormal and idempotent ($\hat{\rho}_n \hat{\rho}_{n'} = \delta_{n, n'} \hat{\rho}_n$), their phase-space Wigner transforms naturally inherit this algebraic relation under the star product. A detailed proof using Gaussian generating functions is provided in Appendix~\ref{app:star_product_proof}.

Projecting the BS equation onto this mesonic basis collapses the differential star product into decoupled, algebraic scalar equations for each mesonic Landau level $n$:
\begin{equation}
    \hat{T}_n(q_\parallel) = 2G + 2G \hat{\Pi}_n(q_\parallel) \hat{T}_n(q_\parallel).
\end{equation}
Therefore, within RPA, the irreducible scattering kernel is driven by the bare contact interaction $2G$ and the projected one-loop polarization tensor $\hat{\Pi}_n(q_\parallel)$. It leads to the exact solution for the full scattering amplitude:
\begin{equation}
    \hat{T}_n(q_\parallel) = \frac{2G}{1 - 2G \hat{\Pi}_n(q_\parallel)}.
\end{equation}
The pole mass $m_{\pi^\pm, n}$ of the charged pion in the $n$-th macroscopic Landau level is determined by the divergence of this scattering amplitude, which corresponds strictly to the zeros of the denominator:
\begin{equation}
    1 - 2G \hat{\Pi}_n(q_\parallel^2 = m_{\pi^\pm, n}^2) = 0.
\end{equation}

To extract the dynamical coefficient $\hat{\Pi}_n(q_\parallel)$ from the unprojected momentum-space tensor $\Pi(q_\parallel, \boldsymbol{q}_\perp)$, we utilize the integral trace property of the Wigner-Weyl transform:
\begin{equation}
    \hat{\Pi}_n(q_\parallel) = \frac{1}{2} \int_0^\infty d\alpha_\pi \, \Pi(q_\parallel, \alpha_\pi) \, \mathcal{P}_n(\alpha_\pi) = \frac{1}{2} \int_0^\infty d\alpha_\pi \, \Pi(q_\parallel, \alpha_\pi) \left[ 2 e^{-\alpha_\pi} (-1)^n L_n(2\alpha_\pi) \right].
    \label{eq:pi_projection}
\end{equation}
This specific projection weight inherently guarantees the correct probability normalization of the mesonic wavefunction. The physical consistency of this formulation is readily verified in the homogeneous weak-field limit. When the macroscopic transverse momentum vanishes ($\boldsymbol{q}_\perp \to \mathbf{0}$, hence $\alpha_\pi \to 0$), the unprojected tensor reduces to its translationally invariant form $\Pi(q_\parallel, \alpha_\pi) \to \Pi_{\mathrm{homo}}(q_\parallel)$. Evaluating the projection for the macroscopic ground state ($n=0$) yields:
\begin{equation}
    \hat{\Pi}_{n=0}(q_\parallel) = \Pi_{\mathrm{homo}}(q_\parallel) \int_0^\infty d\alpha_\pi \, e^{-\alpha_\pi} = \Pi_{\mathrm{homo}}(q_\parallel),
\end{equation}
where the projected amplitude smoothly and analytically recovers the standard vacuum RPA polarization tensor.

The primary computational task is therefore to evaluate the unprojected $\Pi(q_\parallel, \boldsymbol{q}_\perp)$ from the trace of two magnetized quark propagators and perform the projection integral in Eq.~(\ref{eq:pi_projection}). Because $|q_u B| \neq |q_d B|$, the substitution will generate asymmetric products of $L_{n_u}(|q_u B|)$ and $L_{n_d}(|q_d B|)$. Integrating these internal functions against the external mesonic state $L_n(|q_\pi B|)$ produces complex generalized triple Laguerre overlap integrals, which we derive in the subsequent sections.

\section{Dynamics of the Charged Pion in a Magnetic Field}
\label{sec:ch2_pion}

We evaluate the unprojected polarization tensor for the charged pion $\pi^+$ by tracing over the $u$ and $\bar{d}$ quark propagators. Let the internal loop momentum be $p$, and the external mesonic momentum be $q = (q_0, \boldsymbol{q}_\perp, q_z)$. It is required to maintain a non-vanishing transverse momentum $\boldsymbol{q}_\perp \neq \mathbf{0}$, as it governs the spatial Moyal star product and the ensuing Laguerre projection. To simplify the longitudinal kinematics without loss of physical insight regarding mass generation, we evaluate the tensor in the longitudinal rest frame by setting $q_z = 0$. The loop momentum routing is defined such that the $u$-quark carries momentum $k = p + q$ and the $d$-quark carries $p$.

To maintain algebraic consistency with the universal gap equation background $J_1$ defined in Section \ref{sec:ch1_gap_ds}, we evaluate the properly scaled unprojected polarization tensor by factoring out the global Dirac scalar factor of $4$:
\begin{equation}
    \Pi_{\pi^+}^{\mathrm{scaled}}(q) = -i N_c \int \frac{d^4p}{(2\pi)^4} \frac{1}{4} \mathrm{Tr}_{\mathrm{D}} \left[ \gamma_5 \tilde{S}_u(k) \gamma_5 \tilde{S}_d(p) \right].
\end{equation}

Because $\mathcal{P}_\pm^d = \mathcal{P}_\mp^u$, the algebraic multiplication of the longitudinal structures generates crossing terms, yielding an inherent parity phase of $(-1)^{n_u+n_d}$ that aligns with the fundamental spatial basis, where $n_u$ and $n_d$ are the internal Landau level indices. Under the chiral flip ($\gamma_5 \gamma_i \gamma_5 = -\gamma_i$), the transverse trace yields $\mathrm{Tr}_{\mathrm{D}}[-\gamma_i \gamma_j] = 4\delta_{ij}$. Combining all the parity signs, the scaled trace evaluates to an additive combination:
\begin{equation}
    \frac{1}{4}\mathrm{Tr}\left[ \gamma_5 \slashed{D}_{n_u}^u(k) \gamma_5 \slashed{D}_{n_d}^d(p) \right] = (-1)^{n_u+n_d} \left[ \frac{1}{2} S_L \cdot \mathcal{F}_{n_u, n_d}^{LL}(\boldsymbol{k}_\perp, \boldsymbol{p}_\perp) + \mathcal{F}_{n_u, n_d}^{TT}(\boldsymbol{k}_\perp, \boldsymbol{p}_\perp) \right],
\end{equation}
where the longitudinal kinematic invariant $S_L = k_\parallel \cdot p_\parallel - M^2$. The corresponding transverse spatial weighting functions are:
\begin{align}
    \mathcal{F}_{n_u, n_d}^{LL} &= L_{n_u}(2\alpha_u) L_{n_d-1}(2\alpha_d) + L_{n_u-1}(2\alpha_u) L_{n_d}(2\alpha_d), \\
    \mathcal{F}_{n_u, n_d}^{TT} &= 4 (\boldsymbol{k}_\perp \cdot \boldsymbol{p}_\perp) L_{n_u-1}^1(2\alpha_u) L_{n_d-1}^1(2\alpha_d),
\end{align}
where $\alpha_u = \boldsymbol{k}_\perp^2 / |q_u B|$ and $\alpha_d = \boldsymbol{p}_\perp^2 / |q_d B|$. Note that the generalized Laguerre polynomial definition inherently enforces $L_{-1}^1(2\alpha) \equiv 0$. This ensures that unphysical spin configurations in the lowest Landau level (LLL) are analytically annihilated, maintaining accurate state counting and numerical stability.

The spatial overlap factors for the macroscopic mesonic ground state ($n=0$, where $L_0(2\alpha_\pi) = 1$) are obtained by integrating the Dirac trace weights over the transverse phase space. The canonical Type I (scalar) and Type II (vector) overlap integrals are:
\begin{align}
    \mathcal{J}_{n_u, n_d, 0}^{(1)} &= \int d^2\boldsymbol{k}_\perp d^2\boldsymbol{p}_\perp \, e^{-\alpha_u - \alpha_d - \alpha_\pi} (-1)^{n_u+n_d} L_{n_u}(2\alpha_u) L_{n_d}(2\alpha_d), \\
    \mathcal{J}_{n_u, n_d, 0}^{(2)} &= \int d^2\boldsymbol{k}_\perp d^2\boldsymbol{p}_\perp \, e^{-\alpha_u - \alpha_d - \alpha_\pi} (\boldsymbol{k}_\perp \cdot \boldsymbol{p}_\perp) (-1)^{n_u+n_d} L_{n_u-1}^1(2\alpha_u) L_{n_d-1}^1(2\alpha_d),
\end{align}
By mapping the internal phase space via $\boldsymbol{k}_\perp = \boldsymbol{p}_\perp + \boldsymbol{q}_\perp$, the differential volume is conserved ($d^2\boldsymbol{q}_\perp d^2\boldsymbol{p}_\perp = d^2\boldsymbol{k}_\perp d^2\boldsymbol{p}_\perp$). As detailed in Appendix~\ref{app:laguerre_recurrence}, we systematically evaluate these integrals by mapping them into parameter space using generating functions to extract finite numerical recurrence relations. In the weak magnetic field regime ($|q_\pi B| \to 0$), the analytical Wigner projection measure asymptotically collapses into a momentum-space Dirac delta function: $\lim_{|q_\pi B| \to 0} [\frac{1}{\pi |q_\pi B|} d^2\boldsymbol{q}_\perp] e^{-\boldsymbol{q}_\perp^2 / |q_\pi B|} = \delta^{(2)}(\boldsymbol{q}_\perp) d^2\boldsymbol{q}_\perp$. This factorization guarantees that the framework perfectly recovers the continuous vacuum polarization tensor, preserving absolute probability conservation.

Applying the Wigner-Weyl Laguerre projection to the lowest macroscopic ground state ($n=0$) yields the dynamical coefficient:
\begin{equation}
    \hat{\Pi}_{n=0}(q_0^2) = \frac{4 i N_c}{\pi |q_\pi B|} \sum_{n_u, n_d = 0}^\infty \int \frac{d\omega dp_z}{(2\pi)^2} \frac{-\frac{1}{2} S_L \left(\mathcal{J}_{n_u, n_d-1, 0}^{(1)} + \mathcal{J}_{n_u-1, n_d, 0}^{(1)}\right) + \mathcal{J}_{n_u, n_d, 0}^{(2)}}{D_u(\omega+q_0) D_d(\omega)}.
\end{equation}
Here, the inverse quark propagators governing the longitudinal energy poles are $D_u(\omega+q_0) = (\omega+q_0)^2 - p_z^2 - M^2 - 2n_u|q_u B|$, $D_d(\omega) = \omega^2 - p_z^2 - M^2 - 2n_d|q_d B|$ and $\omega=p_0$ for convenience. 

We isolate the vacuum condensate background from the dynamic mesonic excitation via a partial fraction decomposition on the longitudinal invariant:
\begin{equation}
    -\frac{S_L}{D_u D_d} = -\left[ \frac{1}{2D_d} + \frac{1}{2D_u} \right] + \frac{q_0^2/2 - n_u|q_u B| - n_d|q_d B| }{D_u D_d}.
\end{equation}
This decomposition uncouples the polarization tensor into a static background and a dynamic kernel: $\hat{\Pi}_{n=0}(q_0^2) = J_1 + J_m(q_0^2)$. 

The single-denominator terms constitute the vacuum condensate background. Because $D_d(\omega)$ depends exclusively on $n_d$, the infinite summation over $n_u$ acts solely on the spatial overlap factor. Utilizing the completeness relation of the Type I overlap integral derived in Appendix~\ref{app:laguerre_recurrence}, we have:
\begin{equation}
    \sum_{n_u=0}^\infty \mathcal{J}_{n_u, n_d, 0}^{(1)} = \frac{\pi^2}{2} |q_\pi B| |q_d B|.
\end{equation}
Substituting this geometric identity back, the overall prefactor evaluates analytically to the charge-dependent factor $i N_c \frac{|q_d B|}{2\pi}$. Accounting for the $(u,d)$-flavor spaces and combining all the factors, the two single-denominator terms exactly recover the vacuum gap equation prefactor $- i N_c \frac{|q_f B|}{4\pi}$, aligning with the $J_1$ background.

The remainder forms the dynamic two-particle scattering kernel $J_m(q_0^2)$, which governs the charged pion mass generation:
\begin{equation}
    J_m(q_0^2) = - \frac{4 i N_c}{\pi |q_\pi B|} \sum_{n_u, n_d = 0}^\infty \int \frac{d\omega dp_z}{(2\pi)^2} \frac{\mathcal{N}_{\pi^+}}{D_u(\omega+q_0) D_d(\omega)},
\end{equation}
where the scalar dynamic mass scale is $R_{\pi^+}^2 = n_u|q_u B| + n_d|q_d B| - q_0^2/2$, and the consolidated transverse spatial weight is:
\begin{equation}
    \mathcal{N}_{\pi^+} = \frac{1}{2} R_{\pi^+}^2 \left(\mathcal{J}_{n_u, n_d-1, 0}^{(1)} + \mathcal{J}_{n_u-1, n_d, 0}^{(1)}\right) - \mathcal{J}_{n_u, n_d, 0}^{(2)}.
\end{equation}

Applying the Matsubara summation formalism ($\omega \to i\omega_m = i(2m+1)\pi T$), the kernel splits into vacuum and medium components ($J_m = J_m^{\mathrm{vac}} + J_m^{\mathrm{med}}$):
\begin{equation}
    J_m^{\mathrm{vac}}(q_0^2) = - \frac{4 N_c}{\pi |q_\pi B|} \sum_{n_u, n_d = 0}^\infty \int_{-\infty}^\infty \frac{dp_z}{2\pi} \frac{\mathcal{N}_{\pi^+}}{2 E_u E_d} \left[ \frac{E_u + E_d}{q_0^2 - (E_u + E_d)^2} \right],
\end{equation}
\begin{align}
    J_m^{\mathrm{med}}(q_0^2, T) = \frac{4 N_c}{\pi |q_\pi B|} \sum_{n_u, n_d = 0}^\infty \int_{-\infty}^\infty \frac{dp_z}{2\pi} \frac{\mathcal{N}_{\pi^+}}{2 E_u E_d} \Bigg[&  \frac{E_u + E_d}{q_0^2 - (E_u + E_d)^2} \left( n_F(E_u) + n_F(E_d) \right) \nonumber \\ 
    & -  \frac{E_u - E_d}{q_0^2 - (E_u - E_d)^2} \left( n_F(E_u) - n_F(E_d) \right) \Bigg].
\end{align}

To manage ultraviolet divergences, we apply a 3D soft cutoff regularization with a smooth damping function $f_\Lambda(x) = [1 + (x/\Lambda^2)^5]^{-1}$. The transverse regularization scale is designated to depend exclusively on the Landau level of a \textit{single} reference constituent quark. By leaving the conjugate constituent's phase space unconstrained by the UV cutoff, the infinite summation over its discrete energy levels cleanly factors out, ensuring the analytical measure annihilation proceeds without truncation artifacts. 

Substituting the chiral symmetry breaking condition $(1 - 2G J_1 = m_0/M)$ into the RPA pole equation manifests Goldstone's theorem: for a finite current quark mass $m_0$, the generalized mass equation reduces to a form dependent only on $J_m$:
\begin{equation}
\frac{m_0}{M} - 2G \left[ J_m^{\mathrm{vac, cutoff}}(m_{\pi^\pm}^2) + J_m^{\mathrm{med}}(m_{\pi^\pm}^2, T) \right] = 0.
\end{equation}

\begin{figure}[htbp]
    \centering
    \includegraphics[width=0.95\linewidth]{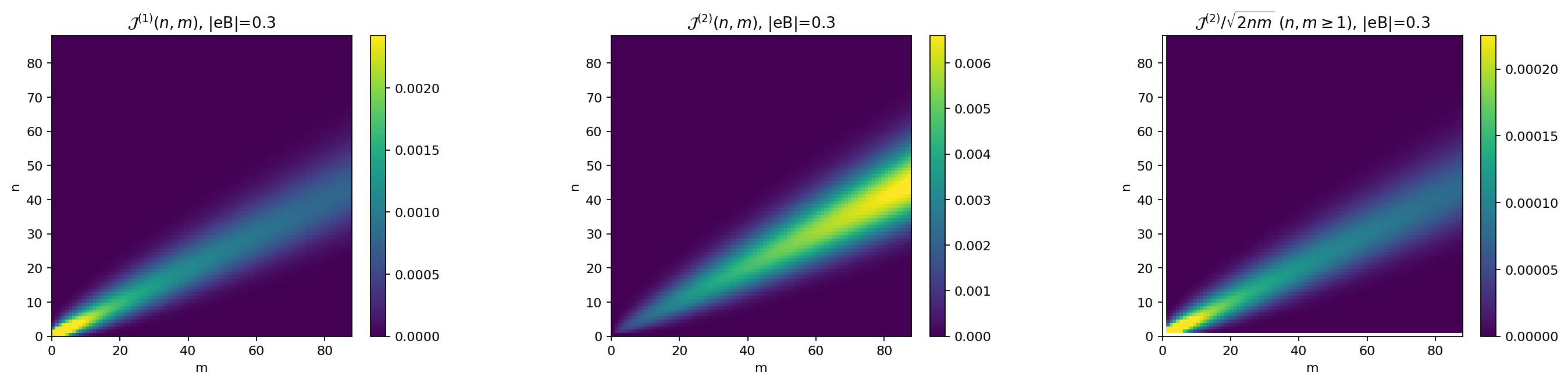}
    \caption{Numerical evaluation of the generalized triple Laguerre overlap integrals in the internal constituent phase space $(n_u, n_d)$ for the macroscopic ground state at $|eB| = 0.3 \text{ GeV}^2$. 
    \textbf{Left:} The Type I scalar overlap $\mathcal{J}^{(1)}(n_u, n_d, 0)$. 
    \textbf{Middle:} The Type II vector overlap $\mathcal{J}^{(2)}(n_u, n_d, 0)$. 
    \textbf{Right:} The normalized Type II overlap. 
    A broad coupling ridge emerges along the trajectory corresponding to the $1:2$ charge ratio, demonstrating the dynamic phase-space matching enforced by the Moyal star product for charged constituents.}
    \label{fig:overlap_nm}
\end{figure}

We first examine the spatial dynamics generated by the asymmetric constituent charges. In Fig.(\ref{fig:overlap_nm}), we present the unprojected spatial overlap integrals in the $(n_u, n_d)$ phase space at $|eB| = 0.3 \text{ GeV}^2$. For a neutral meson, identical charge scalings enforce a Kronecker-delta orthogonality ($n_u = n_d$). In contrast, the absolute constituent charges of the charged pion satisfy a $2:1$ ratio ($|q_u| = 2|q_d|$). As shown in the left and middle panels, the Moyal star product dynamically generates a localized strong-coupling ridge along the kinematic trajectory $n_d \approx 2n_u$, corresponding to the geometric matching of their transverse spatial dispersions. The right panel shows the dimensionally normalized Type II vector overlap, confirming its function as a transverse kinetic energy extraction operator that projects out the scale $\frac{1}{2}\mathcal{C}_{\mathrm{norm}} \cdot (n_f |q_f B|)$.

\begin{figure}[htbp]
    \centering
    \includegraphics[width=0.48\linewidth]{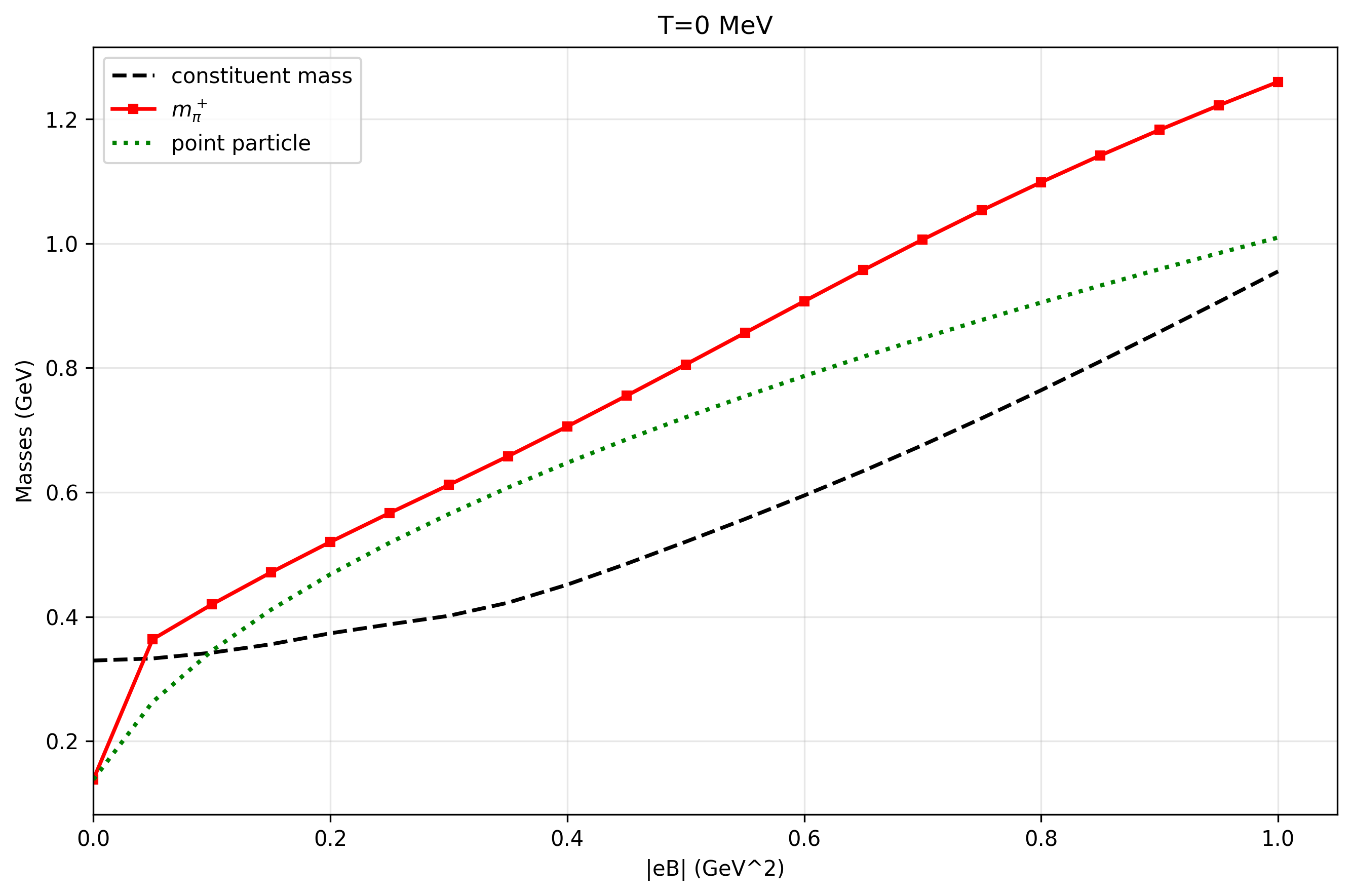}\hfill
    \includegraphics[width=0.48\linewidth]{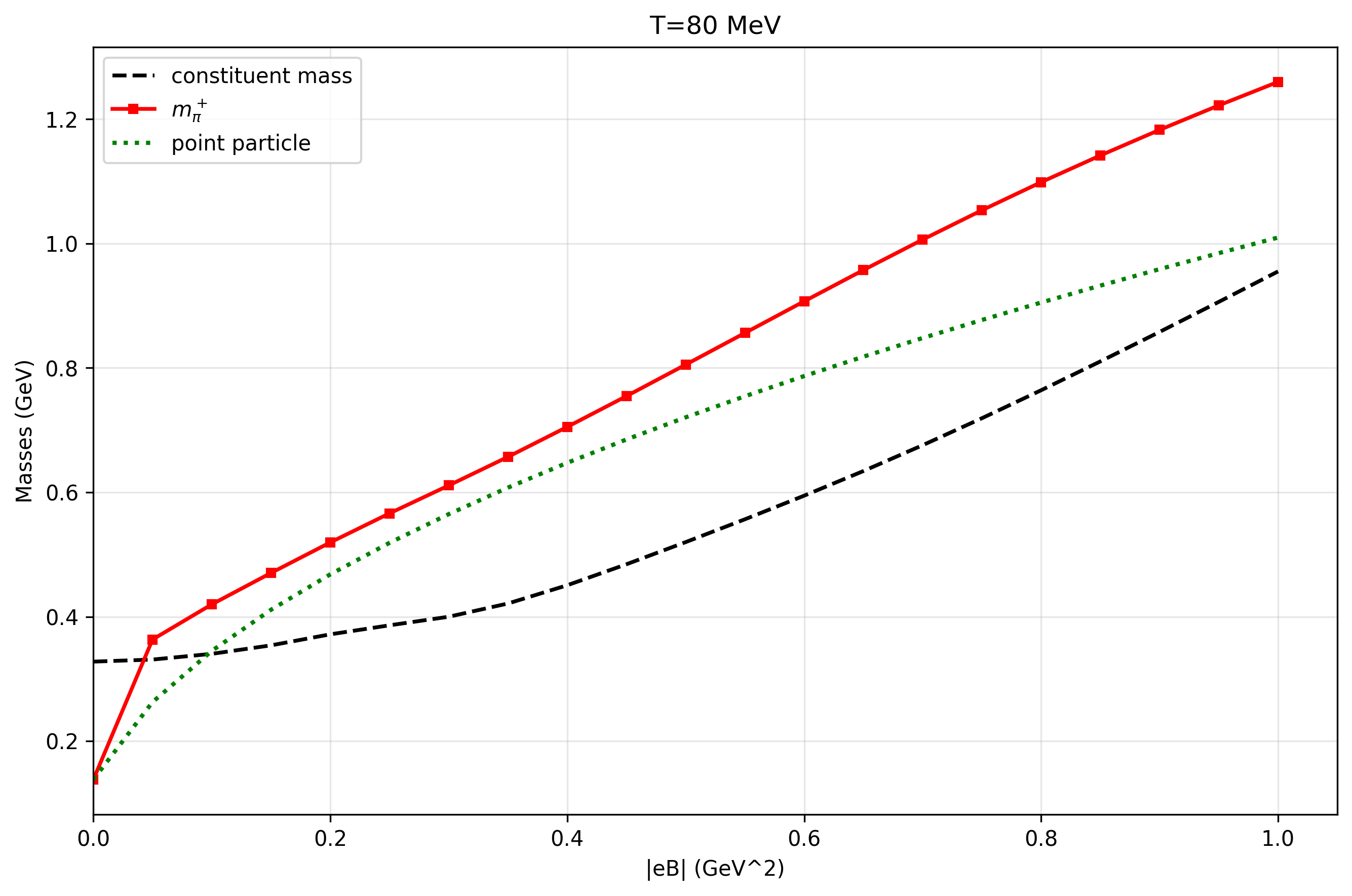}
    \caption{The constituent quark mass $M$ (black dashed line) and the charged pion mass $m_{\pi^+}$ (red line) as a function of the magnetic field $|eB|$ at $T=0$ MeV (left) and $T=80$ MeV (right). The green dotted line represents the point-particle kinematic limit $\sqrt{m_\pi(0)^2 + |eB|}$. The composite pion mass tracks this zero-point drift due to the gap equation cancellation.}
    \label{fig:pion_mass_eB}
\end{figure}

We numerically evaluate the physical mass of the charged pion $\pi^+$. The NJL parameters are fixed at $m_0 = 0.0052 \text{ GeV}$, $\Lambda = 0.6213 \text{ GeV}$, and $G = 5.1881 \text{ GeV}^{-2}$, targeting $m_\pi \approx 138 \text{ MeV}$ and $\langle \bar{u}u \rangle^{1/3} \approx -250 \text{ MeV}$ in the vacuum limit \cite{Coppola:2018vkw, Coppola:2019uyr, Li:2020hlp, Li:2023rsy}. In Fig.(\ref{fig:pion_mass_eB}), we present the constituent quark mass $M$ and the pion mass $m_{\pi^+}$ as functions of the magnetic field at $T=0$ MeV and $T=80$ MeV. The macroscopic mass evolution of the $\pi^+$ rigorously tracks the kinematic zero-point energy drift, $m_{\pi^\pm}^2 \approx m_\pi(0)^2 + |eB|$. Because the dynamically generated vacuum condensate background $J_1$ is canceled exactly by the scalar interaction term via the gap equation, the $\pi^+$ meson remains protected by the generalized Goldstone theorem. The minor numerical non-smoothness near the weak-field limit ($|eB| < 0.05 \text{ GeV}^2$) is a computational artifact arising from evaluating the soft-cutoff regularizer in the increasingly dense discrete continuum.

\begin{figure}[htbp]
    \centering
    \includegraphics[width=0.7\linewidth]{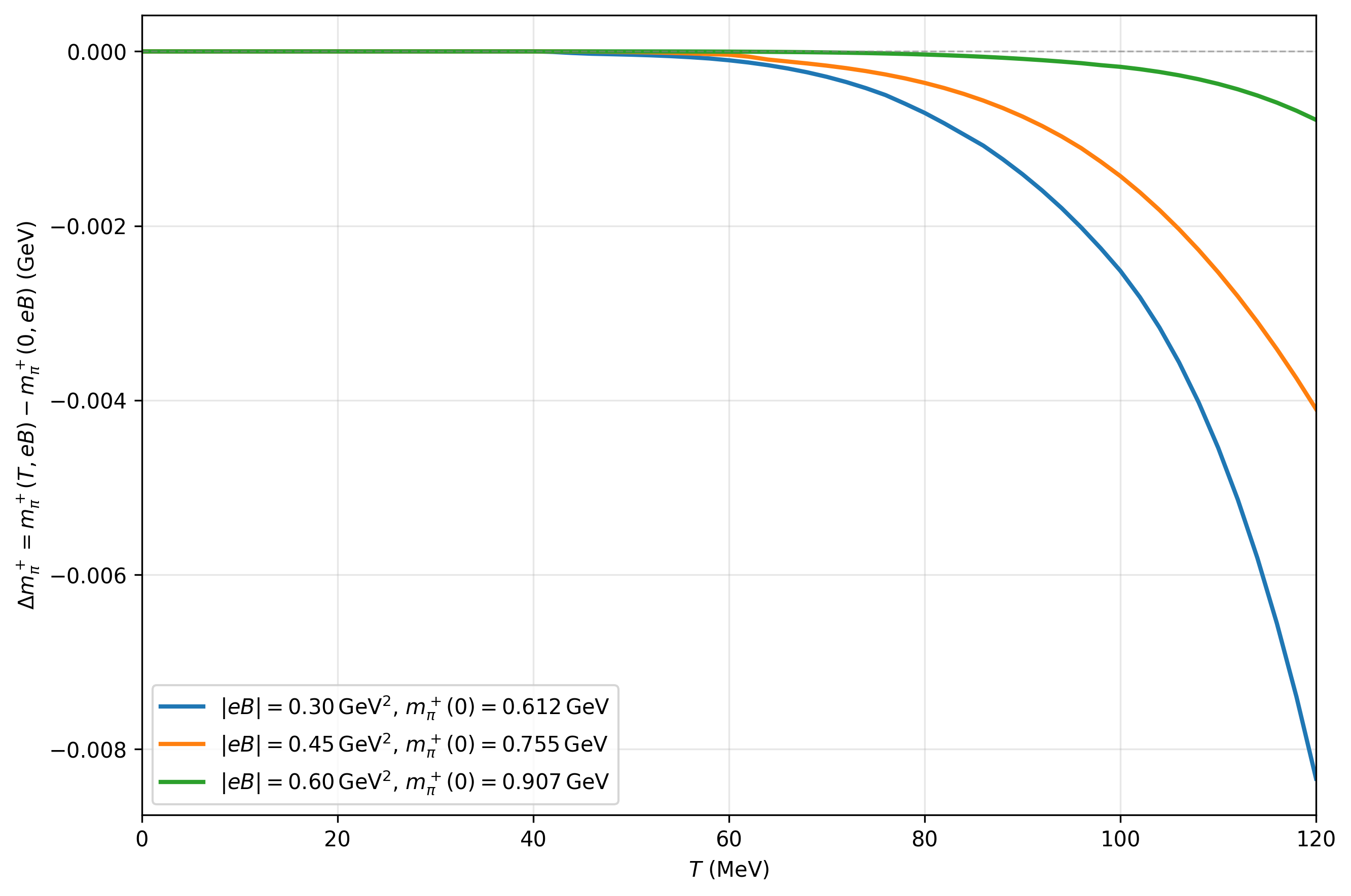}
    \caption{The temperature-induced mass shift $\Delta m_{\pi}^+ = m_{\pi}^+(T, eB) - m_{\pi}^+(0, eB)$ of the charged pion at fixed magnetic fields $|eB| = 0.30, 0.45,$ and $0.60 \text{ GeV}^2$. The downward trend is driven by Pauli blocking and thermal scattering within the finite-temperature Matsubara summations.}
    \label{fig:pion_delta_mass_T}
\end{figure}

Fig.(\ref{fig:pion_delta_mass_T}) illustrates the thermal mass shift $\Delta m_{\pi}^+ = m_{\pi}^+(T, eB) - m_{\pi}^+(0, eB)$. The monotonic suppression of the pion mass as temperature rises is dynamically generated by the Fermi-Dirac distributions within the medium kernel $J_m^{\mathrm{med}}$. Thermal scattering and Pauli blocking effectively weaken the quark-antiquark binding interaction, pushing the pole mass downward prior to the critical temperature of chiral symmetry restoration.

\section{Dynamics of the Charged Rho Meson in a Magnetic Field}
\label{sec:ch3_rho}

The algebraic diagonalization achieved for the pseudoscalar $\pi^+$ meson via the Moyal star product provides a universal framework for charged composite states. We now extend this formalism to the charged vector meson $\rho^+$, which introduces Lorentz tensor structures and spin-magnetic couplings.

The $\rho^+$ meson shares the same valence quark content ($u\bar{d}$) and net electric charge ($q_\rho = e$) as the $\pi^+$ meson. The irreducible quark-antiquark polarization kernel for the vector channel is given by:
\begin{equation}
    K^{\mu\nu}(x, y) = 2G_V \Pi^{\mu\nu}(x, y) = 2i G_V N_c \mathrm{Tr}_{\mathrm{D}}\left[ \gamma^\mu S_u(x, y) \gamma^\nu S_d(y, x) \right],
\end{equation}
where $G_V$ is the vector four-fermion coupling constant. This kernel accumulates the identical macroscopic Schwinger phase: $K^{\mu\nu}(x, y) = e^{i\Phi_\rho(x,y)} \tilde{K}^{\mu\nu}(x-y)$. Factoring out this phase and performing the Fourier transform, the convolution translates into the tensor star product:
\begin{equation}
    \tilde{T}^{\mu\nu}(q) = 2 G_V + \tilde{K}^{\mu\alpha}(q) \star \tilde{T}_\alpha^{\nu}(q).
\end{equation}

To diagonalize this equation, the projection basis must separate the spatial phase space and the Lorentz spin structure. The background magnetic field $\boldsymbol{B} = B\hat{z}$ breaks the $SO(1,3)$ Lorentz symmetry down to $SO(1,1)_\parallel \times SO(2)_\perp$, splitting the physical spin-1 states into three polarization modes governed by the spin projection $s_z$.

We define the orthonormal polarization vectors for the three states $\lambda \in \{+, -, \parallel\}$ \cite{Hidaka:2012mz, Liu:2014uwa, Ghosh:2016evc}:
\begin{align}
    \epsilon_+^\mu &= \frac{1}{\sqrt{2}}(0, 1, i, 0) \quad &(s_z = +1, \text{ Right-handed transverse}), \\
    \epsilon_-^\mu &= \frac{1}{\sqrt{2}}(0, 1, -i, 0) \quad &(s_z = -1, \text{ Left-handed transverse}), \\
    \epsilon_\parallel^\mu &= (0, 0, 0, 1) \quad &(s_z = 0, \text{ Longitudinal}).
\end{align}
These construct the spin projection tensors $P_\lambda^{\mu\nu} = (\epsilon_\lambda^\mu)^* \epsilon_\lambda^\nu$. The complete eigenbasis for the vector meson is the direct product of the macroscopic Laguerre functions $\mathcal{P}_n(\alpha_\rho)$ with $\alpha_\rho \equiv \boldsymbol{q}_\perp^2 / |q_\rho B|$ and the spin projectors $P_\lambda^{\mu\nu}$. Because the spatial star product commutes with constant Lorentz tensors, the projection algebra factorizes:
\begin{equation}
    \left[ P_\lambda^{\mu\alpha} \mathcal{P}_n(\alpha_\rho) \right] \star \left[ P_{\lambda' \alpha}^{\nu} \mathcal{P}_{n'}(\alpha_\rho) \right] = \delta_{n, n'} \delta_{\lambda, \lambda'} \left[ P_\lambda^{\mu\nu} \mathcal{P}_n(\alpha_\rho) \right].
\end{equation}

Projecting the tensor BS equation onto this basis decouples it into three independent scalar equations. The projected dynamical coefficient for each Landau level $n$ and spin state $\lambda$ is:
\begin{equation}
    \hat{\Pi}_{n, \lambda}(q_\parallel) = \int_0^\infty d\alpha_\rho \left[ (\epsilon_\lambda^\mu)^* \Pi_{\mu\nu}(q_\parallel, \alpha_\rho) \epsilon_\lambda^\nu \right] \left[ e^{-\alpha_\rho} (-1)^n L_n(2\alpha_\rho) \right].
\end{equation}
Consequently, the physical masses $m_{\rho^+, n}(\lambda)$ of the charged vector meson are determined by three generalized RPA pole equations:
\begin{equation}
    1 - 2G_V \hat{\Pi}_{n, \lambda}\left(q_\parallel^2 = m_{\rho^+, n}^2(\lambda)\right) = 0 \quad \text{for } \lambda \in \{+, -, \parallel\}.
\end{equation}

The unprojected tensor for a specific physical state $\lambda$ is strictly defined as:
\begin{equation}
    \Pi_{\rho^+}(\lambda) = i N_c \int \frac{d^4p}{(2\pi)^4} \mathrm{Tr}_{\mathrm{D}} \left[ (\gamma \cdot \epsilon_\lambda^*) \tilde{S}_u(k) (\gamma \cdot \epsilon_\lambda) \tilde{S}_d(p) \right].
\end{equation}

For the longitudinal mode ($\lambda = \parallel$), the vertex $\gamma^3$ commutes with the transverse spin projectors $\mathcal{P}_\pm$. Similar to the pseudoscalar channel, this forces a chiral mismatch between the oppositely charged $u$ and $\bar{d}$ quarks, generating misaligned Laguerre pairings. For the transverse modes ($\lambda = \pm$), the chiral vertices ($\gamma^\pm=(\gamma^1\pm i \gamma^2)/2$) dynamically flip the spin projectors ($\gamma^\pm \mathcal{P}_\mp = \mathcal{P}_\pm \gamma^\pm$). This compensates for the underlying charge asymmetry of the $u\bar{d}$ pair, resulting in strictly aligned Laguerre configurations with uniform parity $(-1)^{n_u+n_d}$. 

These chiral vertices also alter the transverse spatial dynamics. By decomposing the transverse momentum operator, $\boldsymbol{\gamma}_\perp \cdot \boldsymbol{k}_\perp = \frac{1}{2}(\gamma^+ k^- + \gamma^- k^+)$, the internal transverse Dirac chain for the right-handed mode ($\lambda = +$) reads:
\begin{equation}
    \gamma^- (\boldsymbol{\gamma}_\perp \cdot \boldsymbol{k}_\perp) \gamma^+ (\boldsymbol{\gamma}_\perp \cdot \boldsymbol{p}_\perp) = \frac{1}{4} \gamma^- (\gamma^+ k^- + \gamma^- k^+) \gamma^+ (\gamma^+ p^- + \gamma^- p^+) \equiv 0.
\end{equation}
An identical algebraic annihilation occurs for the left-handed mode ($\lambda = -$). Physically, this demonstrates that fully polarized transverse vector modes ($|s_z| = 1$) cannot couple to the orbital transverse momentum fluctuations of the internal quark pair, as their chiral spin channels are maximally saturated. Consequently, the transverse momentum trace rigorously vanishes, suppressing any Type II spatial overlap for these two modes. Finally, the scaled unprojected traces evaluate to:
\begin{align}
    \frac{1}{4}\mathrm{Tr}_{\mathrm{Dirac}}[\lambda = \parallel] &= (-1)^{n_u+n_d-1} \left[ \frac{1}{2} S_L^\parallel \cdot \mathcal{F}_{n_u, n_d}^{LL, \parallel}(\boldsymbol{k}_\perp, \boldsymbol{p}_\perp) + \mathcal{F}_{n_u, n_d}^{TT, \parallel}(\boldsymbol{k}_\perp, \boldsymbol{p}_\perp) \right], \\
    \frac{1}{4}\mathrm{Tr}_{\mathrm{Dirac}}[\lambda = \pm] &= (-1)^{n_u+n_d} \left[ S_L^\pm \cdot \mathcal{F}_{n_u, n_d}^{LL, \pm}(\boldsymbol{k}_\perp, \boldsymbol{p}_\perp) \right].
\end{align}
The longitudinal invariants are $S_L^\parallel = k_0 p_0 + k_3 p_3 - M^2$ and $S_L^\pm = k_\parallel \cdot p_\parallel - M^2$. The spatial weighting functions separate according to the chiral algebra:
\begin{align}
    \mathcal{F}_{n_u, n_d}^{LL, \parallel} &= L_{n_u}(2\alpha_u) L_{n_d-1}(2\alpha_d) + L_{n_u-1}(2\alpha_u) L_{n_d}(2\alpha_d), \\
    \mathcal{F}_{n_u, n_d}^{TT, \parallel} &= 4 (\boldsymbol{k}_\perp \cdot \boldsymbol{p}_\perp) L_{n_u-1}^1(2\alpha_u) L_{n_d-1}^1(2\alpha_d), \\
    \mathcal{F}_{n_u, n_d}^{LL, +} &= L_{n_u}(2\alpha_u) L_{n_d}(2\alpha_d), \quad \mathcal{F}_{n_u, n_d}^{LL, -} = L_{n_u-1}(2\alpha_u) L_{n_d-1}(2\alpha_d).
\end{align}

Following the Laguerre projection onto the mesonic ground state ($n=0$, $q_3=0$), the spatial integrations strictly yield the canonical overlaps. For $\lambda = \parallel$:
\begin{equation}
    \mathcal{J}_{n_u, n_d, 0}^{(1), \parallel} = \mathcal{J}_{n_u, n_d-1, 0}^{(1)} + \mathcal{J}_{n_u-1, n_d, 0}^{(1)}, \quad \mathcal{J}_{n_u, n_d, 0}^{(2), \parallel} = \mathcal{J}_{n_u, n_d, 0}^{(2)}.
\end{equation}
For $\lambda = \pm$, the integration isolates only the scalar overlaps:
\begin{equation}
    \mathcal{J}_{n_u, n_d, 0}^{(1), +} = \mathcal{J}_{n_u, n_d, 0}^{(1)}, \quad \mathcal{J}_{n_u, n_d, 0}^{(1), -} = \mathcal{J}_{n_u-1, n_d-1, 0}^{(1)}.
\end{equation}
By performing the algebraic partial fraction decomposition, we separate the static vacuum background from the dynamic scattering kernel. And the projected tensors decouple as $\hat{\Pi}_{\pm}(q_0^2) = J_1 + J_{m, \pm}(q_0^2)$ and $\hat{\Pi}_{\parallel}(q_0^2) = J_1 + J_{m, \parallel}(q_0^2)$. 

Applying the Matsubara summation, the dynamic kernels separate into vacuum and medium components:
\begin{equation}
    J_{m, \lambda}^{\mathrm{vac}}(q_0^2) = - \frac{4 N_c}{\pi |q_\rho B|} \sum_{n_u, n_d = 0}^\infty \int_{-\infty}^\infty \frac{dp_z}{2\pi} \frac{\mathcal{N}_\lambda}{2 E_u E_d} \left[ \frac{E_u + E_d}{q_0^2 - (E_u + E_d)^2} \right],
\end{equation}
\begin{align}
    J_{m, \lambda}^{\mathrm{med}}(q_0^2, T) = \frac{4 N_c}{\pi |q_\rho B|} \sum_{n_u, n_d = 0}^\infty \int_{-\infty}^\infty \frac{dp_z}{2\pi} \frac{\mathcal{N}_\lambda}{2 E_u E_d} \Bigg[ & \frac{E_u + E_d}{q_0^2 - (E_u + E_d)^2} \left( n_F(E_u) + n_F(E_d) \right) \nonumber \\
    - & \frac{E_u - E_d}{q_0^2 - (E_u - E_d)^2} \left( n_F(E_u) - n_F(E_d) \right) \Bigg],
\end{align}
where the consolidated transverse spatial weights are:
\begin{align}
    \mathcal{N}_\parallel &= \frac{1}{2} R_\parallel^2(q_0^2, p_z) \left(\mathcal{J}_{n_u, n_d-1, 0}^{(1)} + \mathcal{J}_{n_u-1, n_d, 0}^{(1)}\right) - \mathcal{J}_{n_u, n_d, 0}^{(2)}, \\
    \mathcal{N}_\pm &= R_\pm^2(q_0^2) \mathcal{J}_{n_u, n_d, 0}^{(1), \pm},
\end{align}
and the dynamic mass scales dictating the thresholds are $R_\pm^2(q_0^2) = n_u|q_u B| + n_d|q_d B| - q_0^2/2$ and $R_\parallel^2(q_0^2, p_z) = R_\pm^2(q_0^2) - 2p_z^2$. The RPA equations governing the masses $m_{\rho^\pm}(\lambda)$ are thus:
\begin{equation}
    1 - 2G_V \left[ J_1 + J_{m, \lambda}(q_0^2 = m_{\rho^\pm, \lambda}^2) \right] = 0.
\end{equation}

A fundamental feature of these equations is that the Zeeman splitting emerges dynamically from microscopic \textit{threshold truncations} enforced by the chiral Dirac algebra, rather than phenomenological pre-assignments. The right-handed state ($\lambda = +$) isolates the aligned pair $L_{n_u} L_{n_d}$, uniquely allowing it to occupy the absolute ground state $(n_u=0, n_d=0)$. Accessing the lowest unshifted kinematic threshold ($2M$), it experiences maximal binding without zero-point magnetic penalties, driving its mass downward ($m^2 - eB$). Conversely, the left-handed state ($\lambda = -$) isolates $L_{n_u-1} L_{n_d-1}$, enforcing a strict truncation ($n_u \ge 1, n_d \ge 1$). The injection of zero-point magnetic energy suppresses the interaction kernel, driving the mass steeply upward ($m^2 + eB$). The longitudinal state ($\lambda = \parallel$) sits structurally between them, requiring exactly one constituent quark to be magnetically excited. 

To regularize the vacuum integrals, we implement the asymmetric 3D soft cutoff established in the pseudoscalar channel, applying the form factor $f_\Lambda$ based exclusively on the longitudinal momentum and Landau level of a single reference constituent quark. Otherwise, a symmetric regularization would entangle the internal phase space and break the non-commutative projector algebra, artificially suppressing the Type II extraction.

\begin{figure}[htbp]
    \centering
    \includegraphics[width=0.7\linewidth]{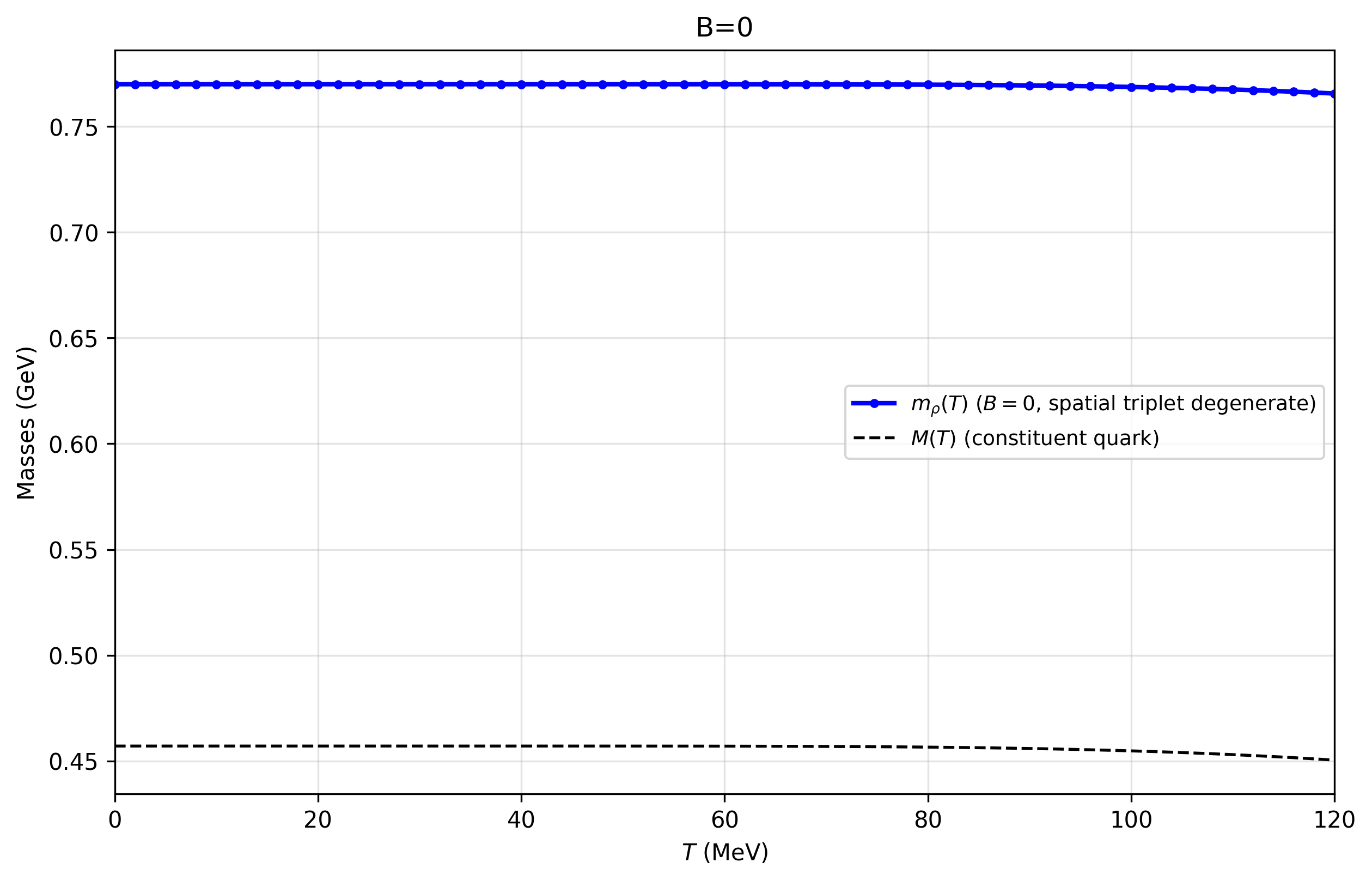}
    \caption{The phenomenological vacuum calibration for the vector channel. The constituent quark mass $M$ (black dashed line) and the unmagnetized $\rho$ meson mass (blue line) are shown as functions of temperature $T$ at $|eB|=0$. The spatial triplet modes are strictly degenerate, fitting to $m_\rho = 0.770 \text{ GeV}$ at $T=0$.}
    \label{fig:rho_vacuum_calibration}
\end{figure}

To numerically evaluate the RPA equations, the parameters are recalibrated for the vector channel. We use the 3D soft-cutoff parameters: $m_0 = 0.005 \text{ GeV}$, $\Lambda = 0.582 \text{ GeV}$, and $G = 2.388 / \Lambda^2 \approx 7.0500 \text{ GeV}^{-2}$. The generated constituent quark mass is $M \approx 0.457 \text{ GeV}$. Targeting $m_\rho = 0.770 \text{ GeV}$ requires $G_V = 5.9432 \text{ GeV}^{-2}$ ($G_V / G \approx 0.843$) \cite{Liu:2014uwa, Carlomagno:2022arc, GomezDumm:2023owj, Coppola:2023mmq}. Fig.(\ref{fig:rho_vacuum_calibration}) illustrates the exact thermal calibration at absolute zero magnetic field, confirming the degenerate triplet states and demonstrating the stability of the baseline prior to symmetry breaking.

\begin{figure}[htbp]
    \centering
    \includegraphics[width=0.48\linewidth]{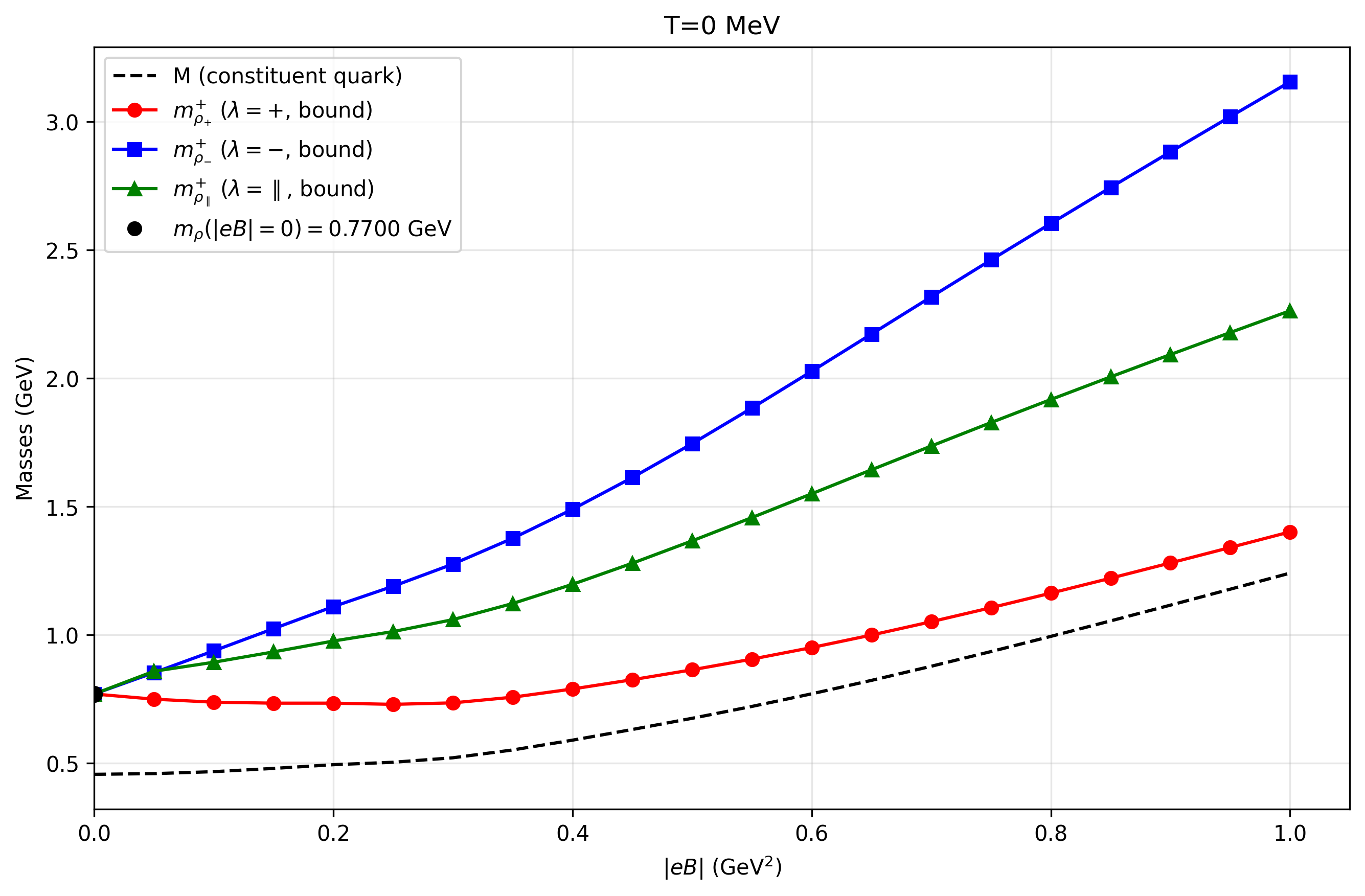}\hfill
    \includegraphics[width=0.48\linewidth]{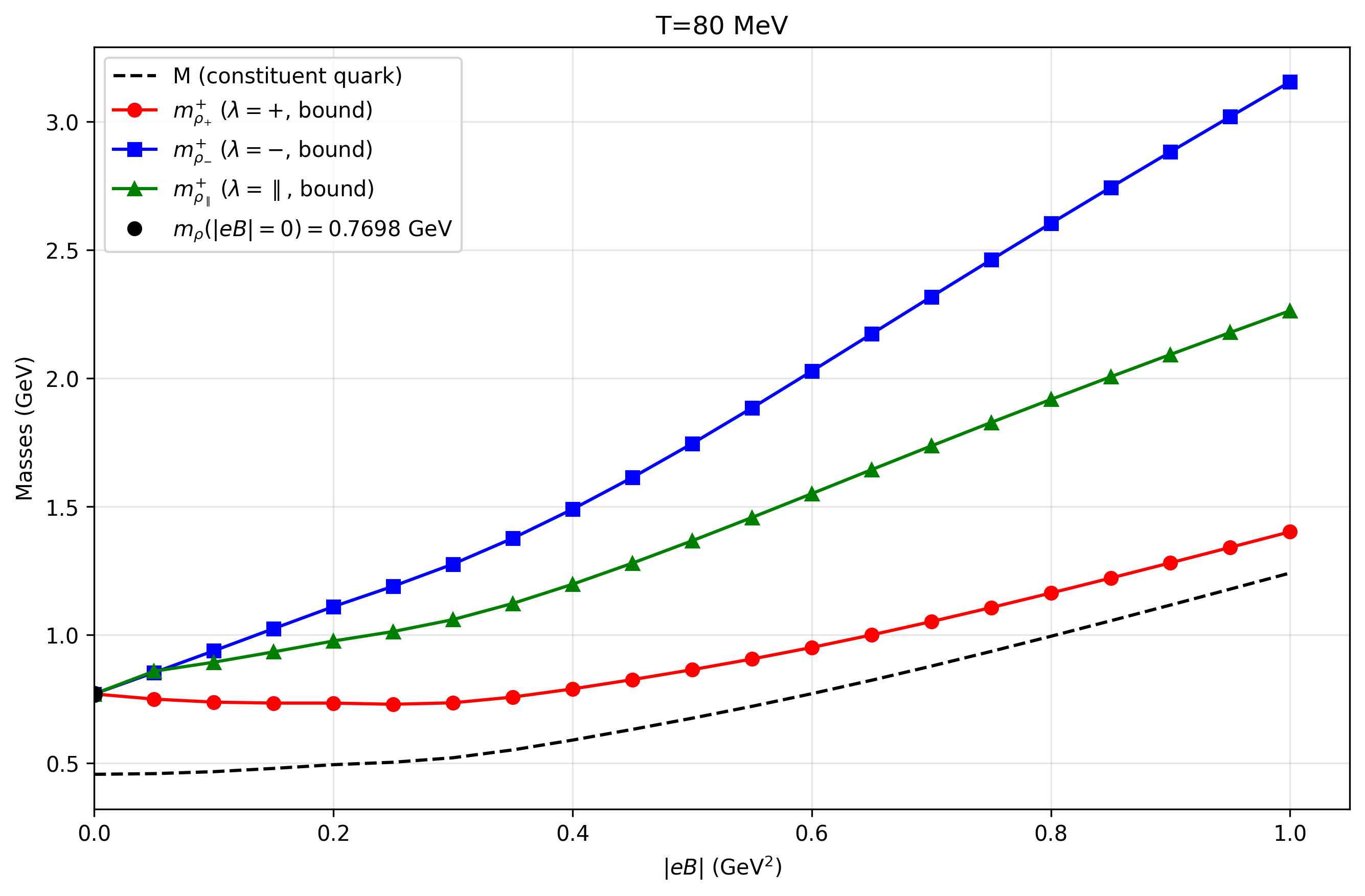}
    \caption{The pole masses of the charged $\rho$ meson triplet states ($\lambda = +, -, \parallel$) as a function of the magnetic field $|eB|$ at $T=0$ MeV (left) and $T=80$ MeV (right). The black dashed line indicates the constituent quark mass $M$. The anticipated VMC for the $\rho^+$ state is dynamically quenched as the MC of the constituent mass overtakes the Zeeman attraction.}
    \label{fig:rho_mass_eB}
\end{figure}

In Fig.(\ref{fig:rho_mass_eB}), we present the numerical solutions for the charged $\rho$ triplet states as functions of $|eB|$ at $T=0$ MeV and $T=80$ MeV. The dynamic Zeeman splitting is unambiguously observed ($m_{\rho^-} > m_{\rho^\parallel} > m_{\rho^+}$). A critical physical behavior emerges regarding the right-handed ($\lambda = +$) state. While early point-particle models predict a tachyonic instability ($m \to 0$) leading to vector meson condensation~\cite{Chernodub:2010qx,Chernodub:2011mc}, our full phase-space evaluation shows an initial dip followed by an arrest and reversal. This is driven by the dynamic competition with magnetic catalysis. The magnetic field rapidly increases the constituent quark mass $M$ (black dashed line), pushing the kinematic threshold ($2M$) upwards. This surge in the threshold fundamentally overtakes the magnetic binding enhancement, quenching the tachyonic instability within this regularized framework.

It is necessary to discuss these findings in the context of recent Lattice QCD evaluations~\cite{DElia:2011koc, Luschevskaya:2014lga, Bali:2017ian, Ding:2020jui, Endrodi:2024cqn}. While our result qualitatively aligns with early non-perturbative calculations indicating quenched VMC, some recent high-precision lattice data suggest a stronger downward trajectory for the $\rho^+$ mass at moderate magnetic fields. This quantitative discrepancy highlights the limitations of the current mean-field approximation. Specifically, effects such as Inverse Magnetic Catalysis (IMC)~\cite{Bruckmann:2013oba, Chao:2013qpa, Li:2016gfn}, where the running coupling constant decreases with $eB$~\cite{Li:2023rsy}, or the explicit incorporation of anomalous magnetic moments (AMM)~\cite{Ferrer:2008dy, Chaudhuri:2019lbw, Xu:2020yag, Xing:2021kbw} for the constituent quarks could weaken the threshold surge and enhance the Zeeman binding, potentially bringing the model closer to recent lattice predictions. These non-trivial magnetic refinements are deferred to subsequent studies.

\begin{figure*}[htbp]
    \centering
    \includegraphics[width=0.32\linewidth]{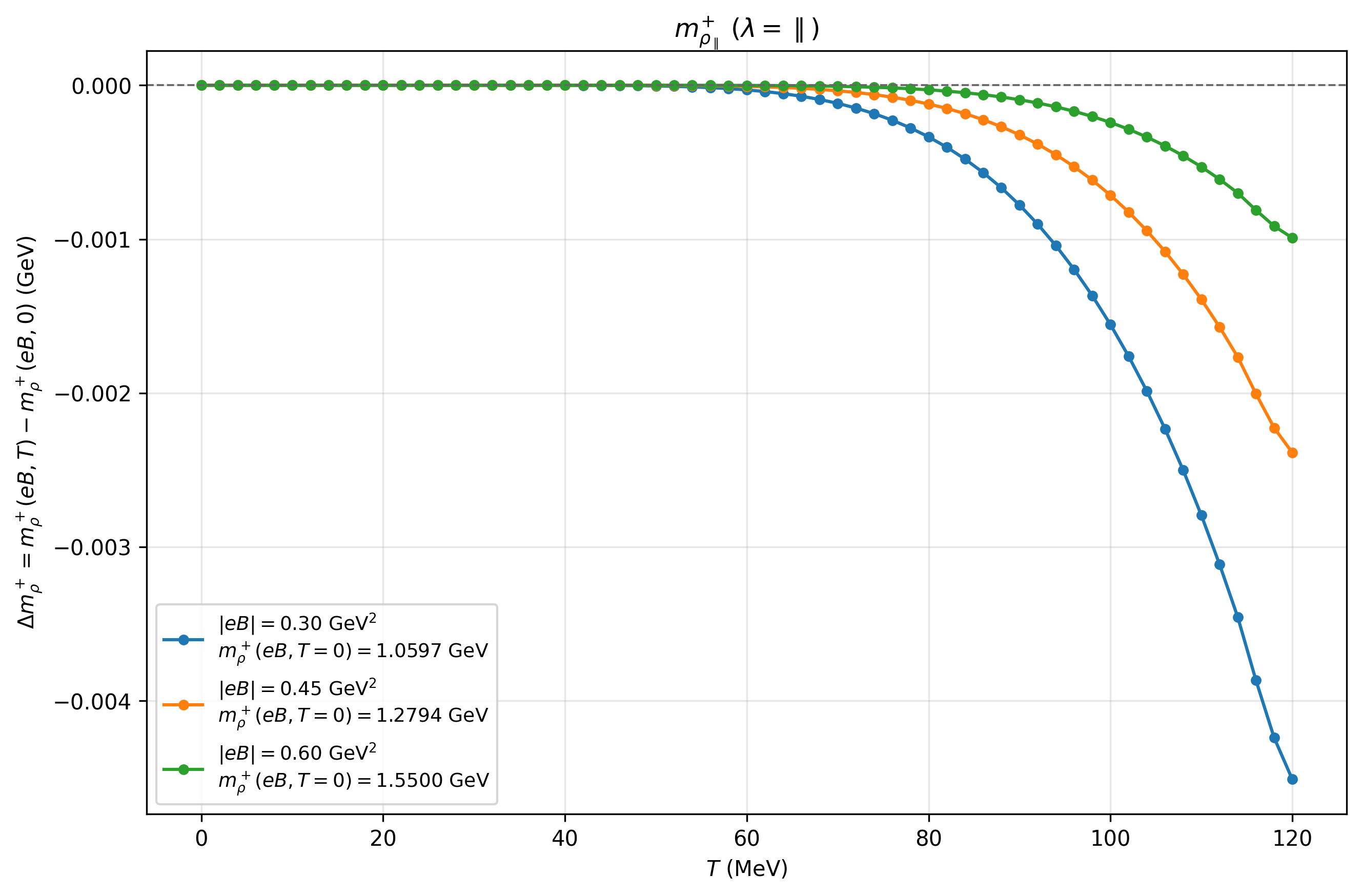}\hfill
    \includegraphics[width=0.32\linewidth]{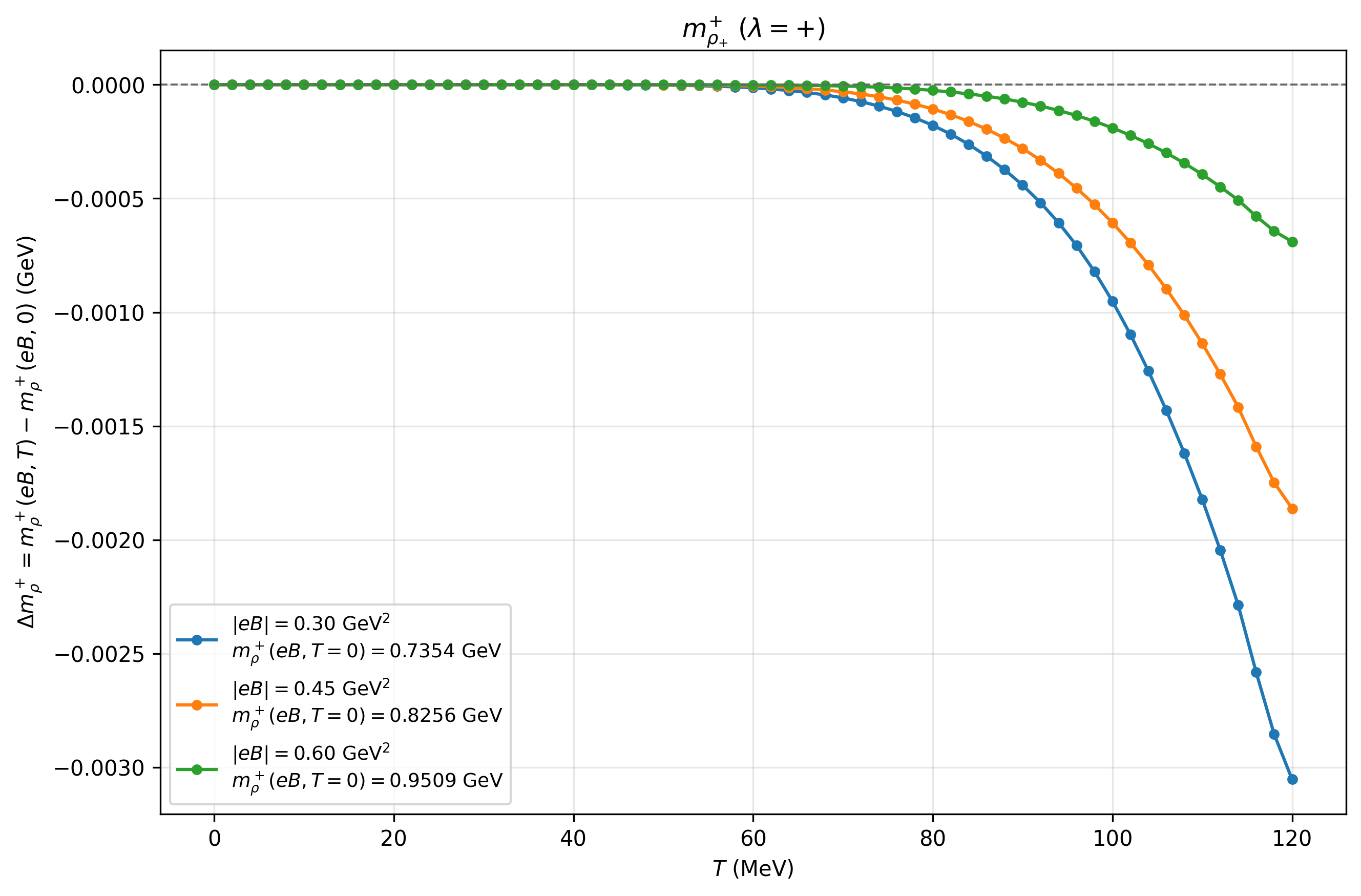}\hfill
    \includegraphics[width=0.32\linewidth]{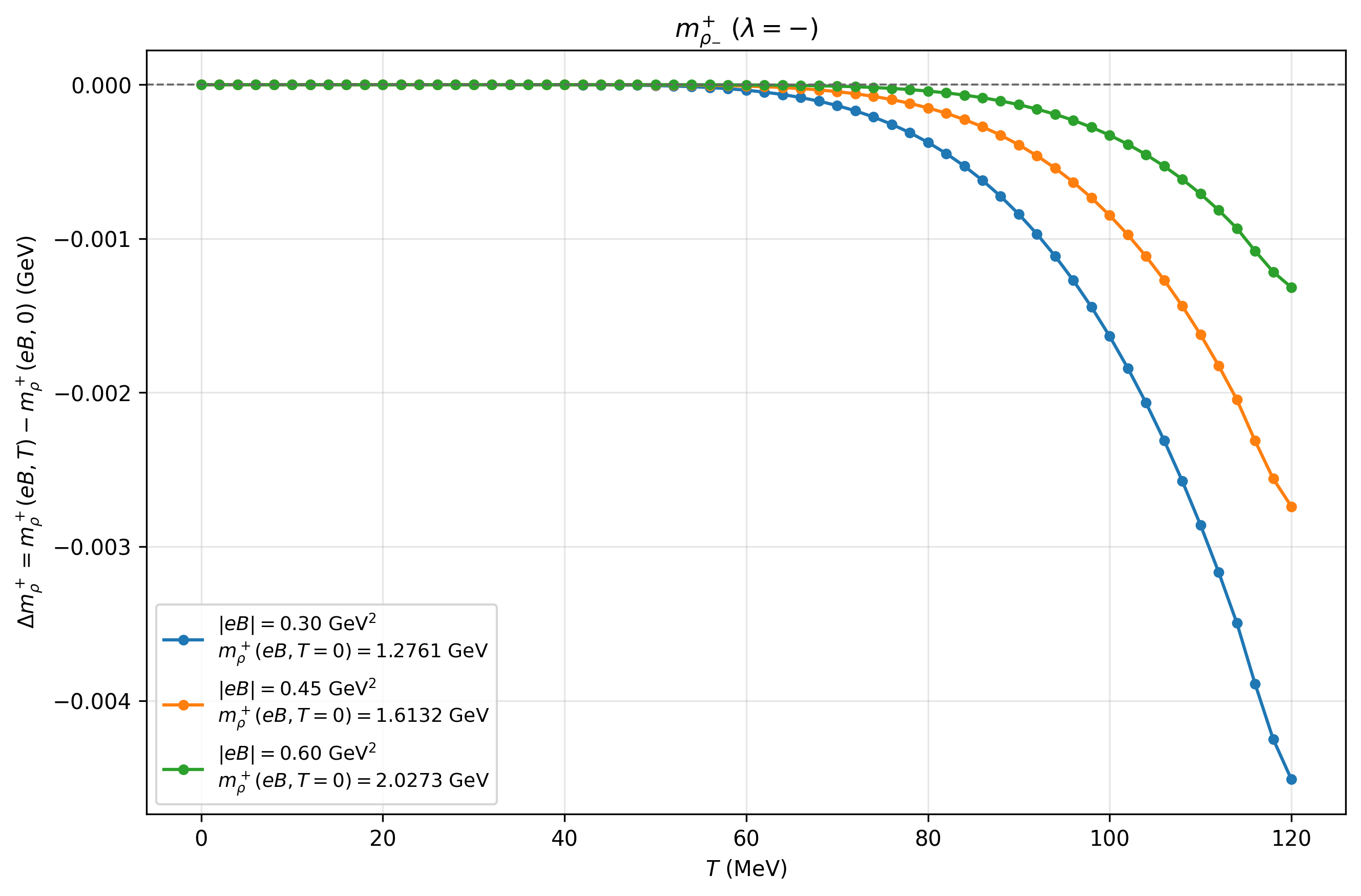}
    \caption{The temperature-induced mass shifts $\Delta m_{\rho} = m_{\rho}(eB, T) - m_{\rho}(eB, 0)$ for the longitudinal ($\lambda = \parallel$, left), right-handed ($\lambda = +$, middle), and left-handed ($\lambda = -$, right) polarization states at fixed magnetic fields. The monotonic thermal suppression is driven by Pauli blocking. All modes maintain bound-state solutions without undergoing Mott dissociation within the evaluated temperature range.}
    \label{fig:rho_delta_m_parallel}
\end{figure*}

Finally, we investigate the thermal evolution of the spin-split states. Fig.(\ref{fig:rho_delta_m_parallel}) displays the temperature-induced mass shifts $\Delta m_{\rho} = m_{\rho}(eB, T) - m_{\rho}(eB, 0)$. Across the computed domain ($T \le 120 \text{ MeV}$), the mass shifts are negative and exhibit monotonic suppression, driven by Pauli blocking within the medium scattering kernel. Importantly, all triplet states remain robustly bound; no Mott dissociation is observed, indicating that the magnetic confinement mechanism is highly resilient against moderate thermal fluctuations prior to the chiral phase transition.

\section{Summary and Conclusions}
\label{sec:summary} 

While mathematically equivalent to the standard Schwinger proper-time integral, our approach offers a distinct physical advantage by explicitly expanding the translationally invariant propagator in the Landau level representation. The traditional proper-time integral effectively integrates out the intermediate discrete states, thereby obscuring the underlying energy spectrum. In contrast, our explicit summation transparently exposes the internal microscopic kinematics and the state-by-state dynamics of the constituent $u$ and $\bar{d}$ quarks, which otherwise remain heavily entangled in formalisms relying on mixed coordinate-momentum representations. To resolve the asymmetric non-local interactions induced by the fractional constituent charges, we formulated an exact non-commutative phase-space framework via the Wigner-Weyl transform and the Moyal star product. This method achieves an algebraic diagonalization of the tensor Bethe–Salpeter integral equations, translating complex spatial convolutions into generalized triple Laguerre overlap integrals. Especially, this explicit phase-space mapping exposes the underlying spatial sum rules. By identifying these analytical sum rules, we constructed an improved, asymmetric ultraviolet truncation scheme that respects the spatial factors. This geometrically consistent regularization circumvents the probability non-conservation introduced by naive cutoffs, guaranteeing mathematically robust and physically transparent bound-state solutions across all magnetic regimes. 

Applying this formalism to the pseudoscalar channel, we analytically verified the cancellation between the spatial sum rules of the RPA kernel and the vacuum gap equation. This identity preserves the generalized Goldstone theorem under magnetic fields. Consequently, the pole mass of the charged pion tracks the kinematic zero-point energy drift of $|eB|$. This behavior validates the consistency of the asymmetric 3D soft-cutoff regularization scheme in the continuum limit. 

In the vector channel, our phase-space evaluation shows that the Zeeman spin-splitting of the $\rho^+$ meson is generated by threshold truncations dictated by the chiral Dirac algebra, rather than being introduced phenomenologically. While the longitudinal ($\lambda = \parallel$) and left-handed ($\lambda = -$) modes acquire zero-point magnetic kinetic energy penalties, the right-handed ($\lambda = +$) state accesses the ground state $(n_u=0, n_d=0)$, leading to an initial Zeeman mass reduction. This mechanism addresses limitations in models that treat the Zeeman shift and constituent mass generation independently. We found that the tachyonic instability of the $\rho^+$ state is quenched by chiral symmetry breaking. Magnetic catalysis drives the constituent quark mass $M$ upwards. Because the mesonic pole is bounded by the continuum threshold ($2M$), the increase of this threshold overtakes the Zeeman attraction, preventing the mass squared from becoming negative and thus avoiding vector meson condensation within this mean-field framework \cite{Ferrer:2008dy, Ding:2025zqu}. 

At finite temperatures, the meson masses exhibit monotonic thermal suppression across all polarization states. This is driven by thermal scattering and Pauli blocking within the Matsubara medium kernel, which reduces the effective quark-antiquark binding. Despite this suppression, all mesonic modes remain bound. The absence of Mott dissociation within the explored temperature range ($T \le 120 \text{ MeV}$) indicates that the magnetic binding remains stable against thermal fluctuations prior to chiral restoration. 

In conclusion, the Moyal star product framework provides a consistent approach for exploring composite states in magnetic fields~\cite{Sheng:2022ssp, Kojo:2026gep}. Future extensions could include the introduction of a finite baryon chemical potential ($\mu_B$)~\cite{Braby:2009dw, Fukushima:2012kc, Li:2016gfn, Avancini:2018svs, Wen:2023qcz, Kojo:2021gvm} or a chiral chemical potential ($\mu_5$)~\cite{Chao:2013qpa, Fukushima:2012kc}. The interplay between finite density and the magnetic field could induce inverse magnetic catalysis, potentially suppressing the $2M$ threshold increase and re-triggering the tachyonic instability \cite{Fukushima:2012kc, Li:2016gfn, Ding:2020inp}. Additionally, adapting this phase-space projection to real-time formalisms will allow for the calculation of spectral functions~\cite{Braby:2010tk,Ghosh:2020qvg,Mei:2026xlj} and transport coefficients of magnetized mesonic matter, relevant for the phenomenological interpretation of heavy-ion collisions \cite{Kharzeev:2015znc, Huang:2015oca, Ghosh:2016evc, Fang:2023bbw}.

\begin{acknowledgments}
This work is supported by the National Natural Science Foundation of China (NSFC) under Grant No.~12365020 and 12505148. 
\end{acknowledgments}

\appendix
\section{Analytical Proof of the Non-Commutative Projector Algebra}
\label{app:star_product_proof}
While the idempotent projector algebra $\mathcal{P}_n(\alpha_\pi) \star \mathcal{P}_{n'}(\alpha_\pi) = \delta_{n, n'} \mathcal{P}_n(\alpha_\pi)$ can be indirectly inferred via the Weyl-Wigner correspondence to the Hilbert space quantum harmonic oscillator, we present here a purely analytical proof operating strictly within the phase-space geometry using generating functions \cite{Moyal:1949sk, Zachos:1999wn, Szabo:2001kg}.

For the Moyal star product between two arbitrary functions $f$ and $g$ in the transverse momentum space, characterized by the non-commutative parameter $\theta = |q_\pi B|$, the infinite-order pseudo-differential operator strictly translates into its well-known non-local integral representation via the Fourier transform. This representation reads~\cite{Langmann:2003cg}:
\begin{equation}
    (f \star g)(\boldsymbol{q}_\perp) = \frac{1}{\pi^2 \theta^2} \int d^2\boldsymbol{u} \int d^2\boldsymbol{v} \, f(\boldsymbol{q}_\perp + \boldsymbol{u}) g(\boldsymbol{q}_\perp + \boldsymbol{v}) \exp\left( \frac{2i}{\theta} (\boldsymbol{u} \times \boldsymbol{v}) \right).
\end{equation}

To evaluate the star product of two concentric Gaussian distributions, we substitute the isotropic functions $f(\boldsymbol{q}_\perp) = \exp(-c_1 \boldsymbol{q}_\perp^2 / \theta) \equiv e^{-c_1 \alpha_\pi}$ and $g(\boldsymbol{q}_\perp) = \exp(-c_2 \boldsymbol{q}_\perp^2 / \theta) \equiv e^{-c_2 \alpha_\pi}$ into the integral representation. The algebraic interplay between the Gaussian exponents and the non-commutative cross-product phase allows for exact analytical integration over the intermediate variables $\boldsymbol{u}$ and $\boldsymbol{v}$. Exploiting the closure of multi-dimensional Gaussian integrals, the convolution rigorously reduces to the fundamental Gaussian star product lemma~\cite{Szabo:2001kg}:
\begin{equation}
    e^{-c_1 \alpha_\pi} \star e^{-c_2 \alpha_\pi} = \frac{1}{1 + c_1 c_2} \exp\left( - \frac{c_1 + c_2}{1 + c_1 c_2} \alpha_\pi \right),
    \label{eq:gaussian_lemma}
\end{equation}
where $c_1$ and $c_2$ are complex constants ensuring convergence.

Recall the explicit definition of the macroscopic mesonic Wigner basis functions~\cite{Berra-Montiel:2024ubb}:
\begin{equation}
    \mathcal{P}_n(\alpha_\pi) = 2 e^{-\alpha_\pi} (-1)^n L_n(2\alpha_\pi).
\end{equation}
To evaluate the star product across all Landau levels simultaneously, we construct the generating function $\mathcal{G}(t, \alpha_\pi)$ for this basis using a formal parameter $t$. Utilizing the standard generalized Laguerre generating series \cite{Gradshteyn:1943cpj}:
\begin{equation}\label{eqn:gen_Lag}
    \sum_{n=0}^\infty L_n(x) z^n = (1-z)^{-1} \exp\left[-\frac{xz}{1-z}\right] \quad \text{for } |z| < 1,
\end{equation}
and substituting $z = -t$ alongside $x = 2\alpha_\pi$, the generating function organizes into a pure Gaussian form:
\begin{align}
    \mathcal{G}(t, \alpha_\pi) &\equiv \sum_{n=0}^\infty t^n \mathcal{P}_n(\alpha_\pi) = 2 e^{-\alpha_\pi} \sum_{n=0}^\infty (-t)^n L_n(2\alpha_\pi) \nonumber \\
    &= \frac{2}{1+t} \exp\left( - \alpha_\pi \frac{1-t}{1+t} \right).
    \label{eq:wigner_generating}
\end{align}

We now evaluate the star product of two independent generating functions, $\mathcal{G}(t_1, \alpha_\pi) \star \mathcal{G}(t_2, \alpha_\pi)$. By defining the rational parameters $c_1 = \frac{1-t_1}{1+t_1}$ and $c_2 = \frac{1-t_2}{1+t_2}$, we directly apply the Gaussian lemma in Eq.~(\ref{eq:gaussian_lemma}). The governing coefficients analytically evaluate to:
\begin{align}
    1 + c_1 c_2 &= \frac{2(1 + t_1 t_2)}{(1+t_1)(1+t_2)}, \\
    c_1 + c_2 &= \frac{2(1 - t_1 t_2)}{(1+t_1)(1+t_2)}.
\end{align}
The combined exponent simplifies into an argument dependent exclusively on the product $t_1 t_2$:
\begin{equation}
    \frac{c_1 + c_2}{1 + c_1 c_2} = \frac{1 - t_1 t_2}{1 + t_1 t_2}.
\end{equation}
Reassembling the prefactors with the exponential kernel yields the strict algebraic identity:
\begin{align}
    \mathcal{G}(t_1, \alpha_\pi) \star \mathcal{G}(t_2, \alpha_\pi) &= \left( \frac{2}{1+t_1} e^{-c_1 \alpha_\pi} \right) \star \left( \frac{2}{1+t_2} e^{-c_2 \alpha_\pi} \right) \nonumber \\
    &= \frac{4}{(1+t_1)(1+t_2)} \left[ \frac{(1+t_1)(1+t_2)}{2(1 + t_1 t_2)} \right] \exp\left( - \alpha_\pi \frac{1 - t_1 t_2}{1 + t_1 t_2} \right) \nonumber \\
    &= \frac{2}{1 + t_1 t_2} \exp\left( - \alpha_\pi \frac{1 - t_1 t_2}{1 + t_1 t_2} \right) \equiv \mathcal{G}(t_1 t_2, \alpha_\pi).
\end{align}

This identity, $\mathcal{G}(t_1, \alpha_\pi) \star \mathcal{G}(t_2, \alpha_\pi) = \mathcal{G}(t_1 t_2, \alpha_\pi)$, demonstrates that the star product maps the phase-space convolution into a simple multiplicative algebra in the parameter space. Expanding both sides via their respective Taylor series:
\begin{align}
    \text{LHS} &= \left( \sum_{n=0}^\infty t_1^n \mathcal{P}_n(\alpha_\pi) \right) \star \left( \sum_{m=0}^\infty t_2^m \mathcal{P}_m(\alpha_\pi) \right) = \sum_{n=0}^\infty \sum_{m=0}^\infty t_1^n t_2^m \left( \mathcal{P}_n(\alpha_\pi) \star \mathcal{P}_m(\alpha_\pi) \right), \\
    \text{RHS} &= \sum_{k=0}^\infty (t_1 t_2)^k \mathcal{P}_k(\alpha_\pi) = \sum_{n=0}^\infty \sum_{m=0}^\infty t_1^n t_2^m \delta_{n,m} \mathcal{P}_n(\alpha_\pi).
\end{align}
Equating the multilinear coefficients of $t_1^n t_2^m$ completes the proof:
\begin{equation}
    \mathcal{P}_n(\alpha_\pi) \star \mathcal{P}_m(\alpha_\pi) = \delta_{n,m} \mathcal{P}_n(\alpha_\pi).
\end{equation}
This establishes the idempotent Wigner projection algebra purely through exact analytical integration, cleanly circumventing the infinite-order pseudo-differential operator via the integral representation.

\section{Triple Laguerre Overlap Integrals and Recurrence Relations}
\label{app:laguerre_recurrence}

Direct integration of Laguerre polynomials in Cartesian coordinates becomes computationally demanding for high Landau levels. To systematically evaluate these integrals, we map the problem into a parameter space using generating functions, which naturally absorb the alternating phases $(-1)^n$ inherent in the magnetized propagators.

\subsection{Type I (Scalar) Overlap Integrals}
For the standard Laguerre polynomials $L_n(x)$ present in the longitudinal trace, we utilize the generating function in Eq.~(\ref{eqn:gen_Lag}). To perform the transverse phase-space integration, we define the composite 2D momentum vector $\mathbf{v} = (\boldsymbol{k}_\perp, \boldsymbol{p}_\perp)^T$. Substituting the explicit definitions of $\alpha_u$, $\alpha_d$, and $\alpha_\pi$ into the generating function product, the exponent organizes into a quadratic form: $-\frac{1}{2} \mathbf{v}^T \mathbf{A}(\vec{t}) \mathbf{v}$, where $\vec{t} \equiv (t_1, t_2, t_3)$ denotes a set of dimensionless auxiliary variables introduced via the generating functions of the macroscopic Laguerre polynomials to decouple the transverse Gaussian integrals. The symmetric $2 \times 2$ phase-space matrix $\mathbf{A}(\vec{t})$ explicitly encodes the charge asymmetry:
\begin{equation}
    \mathbf{A}(t_1, t_2, t_3) = 2
    \begin{pmatrix}
        \frac{1}{|q_u B|(1+t_1)} + \frac{1}{|q_\pi B|(1+t_3)} & -\frac{1}{|q_\pi B|(1+t_3)} \\
        -\frac{1}{|q_\pi B|(1+t_3)} & \frac{1}{|q_d B|(1+t_2)} + \frac{1}{|q_\pi B|(1+t_3)}
    \end{pmatrix}.
\end{equation}

Because the Gaussian integral evaluates to $[\det \mathbf{A}]^{-1}$ for four real variables, the master generating function $\mathcal{G}^{(1)}(t_1, t_2, t_3)$ for the Type I integrals $\mathcal{J}_{n_u, n_d, n}^{(1)}$ evaluates to a rational algebraic fraction:
\begin{equation}
    \mathcal{G}^{(1)}(t_1, t_2, t_3) = \frac{\pi^2 |q_u B| |q_d B|}{(1+t_1)(1+t_2)(1+t_3)} \frac{1}{\det \mathbf{A}(t_1, t_2, t_3)}.
\end{equation}

The off-diagonal cross-coupling terms inherent in the spatial coordinates square out and cancel during the determinant evaluation over the 2D symmetric phase space. Consequently, the determinant depends exclusively on rational powers of the charge ratios $r_u = |q_u / q_\pi|$ and $r_d = |q_d / q_\pi|$. With $r_u + r_d = 1$, it takes the exact form:
\begin{align}
    \det \mathbf{A}(\vec{t}) = \Delta_0 + \Delta_1 t_1 + \Delta_2 t_2 + \Delta_3 t_3 + \Delta_{12} t_1 t_2 + \Delta_{13} t_1 t_3 + \Delta_{23} t_2 t_3 + \Delta_{123} t_1 t_2 t_3,
\end{align}
where the rational coefficients are:
\begin{align}
    \Delta_0 &= 4, \quad \Delta_1 = -4(r_u - r_d), \quad \Delta_2 = 4(r_u - r_d), \quad \Delta_3 = 0, \nonumber \\
    \Delta_{12} &= -4(r_u - r_d)^2, \quad \Delta_{13} = 4, \quad \Delta_{23} = 4, \quad \Delta_{123} = 0.
\end{align}

To extract a robust numerical recurrence relation, we define the master polynomial $\mathcal{P}^{(1)}(\vec{t}) \equiv (1+t_1)(1+t_2)(1+t_3) \det \mathbf{A}(\vec{t})$. The generating function satisfies the identity:
\begin{equation}
    \mathcal{P}^{(1)}(t_1, t_2, t_3) \mathcal{G}^{(1)}(t_1, t_2, t_3) = \pi^2 |q_u B| |q_d B|.
\end{equation}
By expanding $\mathcal{P}^{(1)}(\vec{t}) = \sum_{i,j,k} p_{i,j,k} t_1^i t_2^j t_3^k$ and matching the power series coefficients for any combined index state $(n_u, n_d, n)$ strictly greater than the origin $(0,0,0)$, the constant term cancels out, dictating the fundamental constant-coefficient ladder recurrence:
\begin{equation}
    \sum_{i,j,k} p_{i,j,k} \mathcal{J}_{n_u-i, n_d-j, n-k}^{(1)} = 0.
\end{equation}
This linear algebraic network requires no index-weighted prefactors, mathematically guaranteeing numerical stability at high Landau levels.

\subsection{Completeness and Probability Conservation}
The physical consistency of the spatial projection framework is verified by evaluating the completeness limit for the macroscopic ground state ($n=0$). Summing over all possible internal states of one constituent quark must analytically recover a constant density of states. 

Recall the integral definition for the canonical Type I overlap in the mesonic ground state ($L_0(2\alpha_\pi) = 1$):
\begin{equation}
    \mathcal{J}_{n_u, n_d, 0}^{(1)} = \int d^2\boldsymbol{k}_\perp d^2\boldsymbol{p}_\perp \, e^{-\alpha_u - \alpha_d - \alpha_\pi} (-1)^{n_u+n_d} L_{n_u}(2\alpha_u) L_{n_d}(2\alpha_d).
\end{equation}
Summing this expression over the $u$-quark Landau levels moves the series inside the integration:
\begin{equation}
    \sum_{n_u=0}^\infty \mathcal{J}_{n_u, n_d, 0}^{(1)} = \int d^2\boldsymbol{k}_\perp d^2\boldsymbol{p}_\perp \, e^{-\alpha_u - \alpha_d - \alpha_\pi} (-1)^{n_d} L_{n_d}(2\alpha_d) \left[ \sum_{n_u=0}^\infty (-1)^{n_u} L_{n_u}(2\alpha_u) \right].
\end{equation}

The evaluation of this infinite sum relies on a sequence of standard analytic reductions. First, applying the completeness relation $\sum_{n_u=0}^\infty (-1)^{n_u} L_{n_u}(2\alpha_u) = \frac{1}{2} e^{\alpha_u}$ entirely eliminates the $u$-quark polynomial structure. With the kinematic relation $\alpha_\pi = (\boldsymbol{k}_\perp - \boldsymbol{p}_\perp)^2 / |q_\pi B|$, the integration over the internal momentum $d^2\boldsymbol{k}_\perp$ factors into a decoupled 2D Gaussian, yielding a geometric factor of $\pi |q_\pi B|$. Finally, the integration over the $d$-quark phase space ($d^2\boldsymbol{p}_\perp = \pi |q_d B| d\alpha_d$) is resolved via the standard Laguerre orthogonality relation $\int_0^\infty e^{-\alpha_d} L_{n_d}(2\alpha_d) d\alpha_d = (-1)^{n_d}$ \cite{Gradshteyn:1943cpj}. This strictly neutralizes the global alternating phase $(-1)^{n_d}$. 

Combining these integrations, the infinite sum smoothly collapses to a purely geometric constant:
\begin{equation}
    \sum_{n_u=0}^\infty \mathcal{J}_{n_u, n_d, 0}^{(1)} = \frac{\pi^2}{2} |q_d B| |q_\pi B| \equiv \mathcal{C}_{\mathrm{norm}}.
\end{equation}
This mathematically proves that $\mathcal{C}_{\mathrm{norm}}$ is a universal constant independent of the initial state $n_d$, ensuring proper normalization without leakage across the asymmetric charge geometries. Notably, this summation identity is inherently symmetric under the flavor exchange $u \leftrightarrow d$, as an equivalent convergence is obtained by summing over the $n_d$ Landau levels for a fixed $n_u$, thereby maintaining the internal consistency of the Wigner-Weyl projection framework.

The Type II vector overlap integral $\mathcal{J}^{(2)}$ encloses the localized extraction of transverse kinetic energy. For the macroscopic ground state, it reads:
\begin{equation}
    \mathcal{J}_{n_u, n_d, 0}^{(2)} = \int d^2\boldsymbol{k}_\perp d^2\boldsymbol{p}_\perp \, e^{-\alpha_u - \alpha_d - \alpha_\pi} (\boldsymbol{k}_\perp \cdot \boldsymbol{p}_\perp) (-1)^{n_u+n_d} L_{n_u-1}^1(2\alpha_u) L_{n_d-1}^1(2\alpha_d).
\end{equation}
For the lowest Landau level ($n_u=0$), the associated Laguerre polynomial $L_{-1}^1 \equiv 0$, rendering the state physically trivial. The infinite summation must strictly commence from $n_u=1$:
\begin{equation}
    \sum_{n_u=1}^\infty \mathcal{J}_{n_u, n_d, 0}^{(2)} = \int d^2\boldsymbol{k}_\perp d^2\boldsymbol{p}_\perp e^{-\alpha_u - \alpha_d - \alpha_\pi} (\boldsymbol{k}_\perp \cdot \boldsymbol{p}_\perp) (-1)^{n_d} L_{n_d-1}^1(2\alpha_d) \left[ \sum_{n_u=1}^\infty (-1)^{n_u} L_{n_u-1}^1(2\alpha_u) \right].
\end{equation}

The bracketed term collapses to $-\frac{1}{4} e^{\alpha_u}$, annihilating the $u$-quark integration measure. Applying the kinematic shift $\boldsymbol{k}_\perp = \boldsymbol{q}_\perp + \boldsymbol{p}_\perp$, the inner product expands to $(\boldsymbol{q}_\perp \cdot \boldsymbol{p}_\perp + \boldsymbol{p}_\perp^2)$. The odd-parity cross term $(\boldsymbol{q}_\perp \cdot \boldsymbol{p}_\perp)$ strictly vanishes upon integration over the symmetric 2D domain. The residual term exclusively isolates the momentum squared of the spectator quark: $\boldsymbol{p}_\perp^2 (\pi |q_\pi B|)$.

Substituting this topological reduction back leaves an integration solely over the $d$-quark configurations:
\begin{equation}
    \sum_{n_u=1}^\infty \mathcal{J}_{n_u, n_d, 0}^{(2)} = \left(-\frac{\pi}{4} |q_\pi B| \right) \int d^2\boldsymbol{p}_\perp \, e^{-\alpha_d} \boldsymbol{p}_\perp^2 (-1)^{n_d} L_{n_d-1}^1(2\alpha_d).
\end{equation}
By invoking the exact differential identity $x L_{n-1}^1(x) = n L_{n-1}(x) - n L_n(x)$ \cite{Gradshteyn:1943cpj}, the residual integral over $x = 2\alpha_d$ resolves to $-n_d (-1)^{n_d}$. This neutralizes both the global alternating phase and the initial negative coefficient, yielding:
\begin{equation}
    \sum_{n_u=1}^\infty \mathcal{J}_{n_u, n_d, 0}^{(2)} = \frac{\pi^2}{4} |q_\pi B| |q_d B| \cdot \left( n_d |q_d B| \right) \equiv \frac{1}{2} \mathcal{C}_{\mathrm{norm}} \cdot \left( n_d |q_d B| \right).
\end{equation}
Unlike the Type I scalar projection that yields a universal geometric constant, the Type II sum rule operates as a transverse kinetic energy projector. It explicitly extracts the intrinsic transverse scale ($n_d |q_d B|$) of the spectator $d$-quark, with the exact same mechanism applying to the $u$-quark.

\subsection{Type II (Vector) Overlaps and Auxiliary Recurrence}
Type II integrals $\mathcal{J}^{(2)}$ involve the bilinear momentum factor $(\boldsymbol{k}_{\perp} \cdot \boldsymbol{p}_{\perp})$. We express this inner product as a symmetric quadratic form over $\mathbf{v}$:
\begin{equation}
    \boldsymbol{k}_\perp \cdot \boldsymbol{p}_\perp = \frac{1}{2} \mathbf{v}^T \mathbf{M} \mathbf{v}, \quad \text{where } \mathbf{M} = \begin{pmatrix} 0 & 1 \\ 1 & 0 \end{pmatrix}.
\end{equation}
According to the Gaussian moment generating theorem, inserting this quadratic form evaluates to $\frac{1}{2}\text{Tr}(\mathbf{M}\mathbf{A}^{-1})$ multiplied by the bare scalar integral. Because $\mathbf{A}^{-1} = \text{adj}(\mathbf{A}) / \det \mathbf{A}$, the raw generating function modifies to:
\begin{equation}
    \mathcal{G}^{(2)}(\vec{t}) = \frac{\pi^2 |q_u B| |q_d B|}{(1+t_1)^2 (1+t_2)^2 (1+t_3)} \frac{\frac{1}{2}\mathrm{Tr}[\mathbf{M}\, \mathrm{adj}(\mathbf{A})]}{[\det \mathbf{A}(\vec{t})]^2}.
\end{equation}

By substituting $\lambda_i = \frac{1-t_i}{1+t_i}$ into $\mathbf{A}$, the off-diagonal coupling yields $\frac{1}{2}\text{Tr}(\mathbf{M}\text{adj}\mathbf{A}) = \frac{2}{|q_\pi B|} \frac{1-t_3}{1+t_3}$. Concurrently, because the $\lambda_3^2$ terms cancel out in $\det \mathbf{A}$, its denominator is strictly $(1+t_1)(1+t_2)(1+t_3)$. The squared determinant inversion thus perfectly annihilates the external measure:
\begin{align}
    \mathcal{G}^{(2)}(\vec{t}) &= \frac{\pi^2 |q_u B| |q_d B|}{(1+t_1)^2 (1+t_2)^2 (1+t_3)} \times \left( \frac{2(1-t_3)}{|q_\pi B| (1+t_3)} \right) \times \frac{(1+t_1)^2 (1+t_2)^2 (1+t_3)^2}{ [\mathcal{P}^{(1)}(\vec{t})]^2 } \nonumber \\
    &= \frac{2\pi^2 |q_u B| |q_d B| / |q_\pi B|}{[\mathcal{P}^{(1)}(\vec{t})]^2} (1 - t_3).
\end{align}
Here, $\mathcal{P}^{(1)}(\vec{t})$ is the cleared determinant polynomial from the Type I framework. The vector coupling polynomial rigorous collapses into a simple first-order binomial dependent exclusively on the external mesonic variable $t_3$:
\begin{equation}
    P_{\mathrm{vec}}(\vec{t}) = 1 - t_3.
\end{equation}

To evaluate $\mathcal{J}^{(2)}$ efficiently, we introduce a set of auxiliary scalar integrals $\mathcal{I}_{m_u, m_d, n}$, which are defined exclusively by the squared denominator $[\mathcal{P}^{(1)}(\vec{t})]^{-2}$ of the generating function. The algebraic structure of this denominator dictates that these auxiliary integrals satisfy a finite homogeneous recurrence relation. By expanding the characteristic polynomial as $\mathcal{P}_{\mathrm{aux}}(\vec{t}) \equiv [\mathcal{P}^{(1)}(\vec{t})]^2 = \sum q_{i,j,k} t_1^i t_2^j t_3^k$, the recurrence is explicitly given by:
\begin{equation}
    \sum_{i,j,k} q_{i,j,k} \mathcal{I}_{m_u-i, m_d-j, n-k} = 0.
\end{equation}

The physical Type II overlaps are reconstructed by applying $P_{\mathrm{vec}}$ onto the auxiliary lattice. Within the parameter space, multiplying the series by $t_3$ corresponds to a backwards discrete index shift on the mesonic Landau level $n$. Introducing the shift operator $\hat{\mathcal{S}}_3 \mathcal{I}_{m_u, m_d, n} = \mathcal{I}_{m_u, m_d, n-1}$, the reconstruction forms a direct difference equation:
\begin{equation}
    \mathcal{J}_{m_u+1, m_d+1, n}^{(2)} \propto P_{\mathrm{vec}}(\hat{\mathcal{S}}_3) \mathcal{I}_{m_u, m_d, n} = \mathcal{I}_{m_u, m_d, n} - \mathcal{I}_{m_u, m_d, n-1}.
\end{equation}
This algebraic decoupling mechanism preserves exact double-precision accuracy and circumvents the numerical instability intrinsic to high-dimensional Laguerre integrations.

\bibliography{reference}

@Article{Kaspi:2017fwg,
  author        = {Kaspi, Victoria M. and Beloborodov, Andrei},
  journal       = {Ann. Rev. Astron. Astrophys.},
  title         = {{Magnetars}},
  year          = {2017},
  pages         = {261--301},
  volume        = {55},
  archiveprefix = {arXiv},
  doi           = {10.1146/annurev-astro-081915-023329},
  eprint        = {1703.00068},
  primaryclass  = {astro-ph.HE},
}

@Article{Andersen:2012dz,
  author        = {Andersen, Jens O.},
  journal       = {Phys. Rev. D},
  title         = {{Thermal pions in a magnetic background}},
  year          = {2012},
  pages         = {025020},
  volume        = {86},
  archiveprefix = {arXiv},
  doi           = {10.1103/PhysRevD.86.025020},
  eprint        = {1202.2051},
  primaryclass  = {hep-ph},
}

@Article{Zhang:2016qrl,
  author        = {Zhang, Rui and Fu, Wei-jie and Liu, Yu-xin},
  journal       = {Eur. Phys. J. C},
  title         = {{Properties of Mesons in a Strong Magnetic Field}},
  year          = {2016},
  number        = {6},
  pages         = {307},
  volume        = {76},
  archiveprefix = {arXiv},
  doi           = {10.1140/epjc/s10052-016-4123-8},
  eprint        = {1604.08888},
  primaryclass  = {hep-ph},
}

@Article{Wang:2017vtn,
  author        = {Wang, Ziyue and Zhuang, Pengfei},
  journal       = {Phys. Rev. D},
  title         = {{Meson properties in magnetized quark matter}},
  year          = {2018},
  number        = {3},
  pages         = {034026},
  volume        = {97},
  archiveprefix = {arXiv},
  doi           = {10.1103/PhysRevD.97.034026},
  eprint        = {1712.00554},
  primaryclass  = {hep-ph},
}

@Article{Liu:2018zag,
  author        = {Liu, Hao and Wang, Xinyang and Yu, Lang and Huang, Mei},
  journal       = {Phys. Rev. D},
  title         = {{Neutral and charged scalar mesons, pseudoscalar mesons, and diquarks in magnetic fields}},
  year          = {2018},
  number        = {7},
  pages         = {076008},
  volume        = {97},
  archiveprefix = {arXiv},
  doi           = {10.1103/PhysRevD.97.076008},
  eprint        = {1801.02174},
  primaryclass  = {hep-ph},
}

@Article{Chaudhuri:2019lbw,
  author        = {Chaudhuri, Nilanjan and Ghosh, Snigdha and Sarkar, Sourav and Roy, Pradip},
  journal       = {Phys. Rev. D},
  title         = {{Effect of the anomalous magnetic moment of quarks on the phase structure and mesonic properties in the NJL model}},
  year          = {2019},
  number        = {11},
  pages         = {116025},
  volume        = {99},
  archiveprefix = {arXiv},
  doi           = {10.1103/PhysRevD.99.116025},
  eprint        = {1907.03990},
  primaryclass  = {nucl-th},
}

@Article{Sheng:2020hge,
  author        = {Sheng, Bingkai and Wang, Yuanyuan and Wang, Xinyang and Yu, Lang},
  journal       = {Phys. Rev. D},
  title         = {{Pole and screening masses of neutral pions in a hot and magnetized medium: A comprehensive study in the Nambu\textendash{}Jona-Lasinio model}},
  year          = {2021},
  number        = {9},
  pages         = {094001},
  volume        = {103},
  archiveprefix = {arXiv},
  doi           = {10.1103/PhysRevD.103.094001},
  eprint        = {2010.05716},
  primaryclass  = {hep-ph},
}

@Article{Agasian:2001ym,
  author        = {Agasian, Nikita O and Shushpanov, Ivan A},
  journal       = {JHEP},
  title         = {{Gell-Mann-Oakes-Renner relation in a magnetic field at finite temperature}},
  year          = {2001},
  pages         = {006},
  volume        = {10},
  archiveprefix = {arXiv},
  doi           = {10.1088/1126-6708/2001/10/006},
  eprint        = {hep-ph/0107128},
}

@Article{Ding:2020jui,
  author        = {Ding, Heng-Tong and Li, Sheng-Tai and Mukherjee, Swagato and Tomiya, Akio and Wang, Xiao-Dan},
  journal       = {PoS},
  title         = {{Meson masses in external magnetic fields with HISQ fermions}},
  year          = {2020},
  pages         = {250},
  volume        = {LATTICE2019},
  archiveprefix = {arXiv},
  doi           = {10.22323/1.363.0250},
  eprint        = {2001.05322},
  primaryclass  = {hep-lat},
}

@Article{Ferrer:2008dy,
  author        = {Ferrer, Efrain J. and de la Incera, Vivian},
  journal       = {Phys. Rev. Lett.},
  title         = {{Dynamically Induced Zeeman Effect in Massless QED}},
  year          = {2009},
  pages         = {050402},
  volume        = {102},
  archiveprefix = {arXiv},
  doi           = {10.1103/PhysRevLett.102.050402},
  eprint        = {0807.4744},
  primaryclass  = {hep-ph},
}

@Article{Schwinger:1951nm,
  author  = {Schwinger, Julian S.},
  journal = {Phys. Rev.},
  title   = {{On gauge invariance and vacuum polarization}},
  year    = {1951},
  pages   = {664--679},
  volume  = {82},
  doi     = {10.1103/PhysRev.82.664},
  editor  = {Milton, K. A.},
}

@Article{DElia:2018xwo,
  author        = {D'Elia, Massimo and Manigrasso, Floriano and Negro, Francesco and Sanfilippo, Francesco},
  journal       = {Phys. Rev. D},
  title         = {{QCD phase diagram in a magnetic background for different values of the pion mass}},
  year          = {2018},
  number        = {5},
  pages         = {054509},
  volume        = {98},
  archiveprefix = {arXiv},
  doi           = {10.1103/PhysRevD.98.054509},
  eprint        = {1808.07008},
  primaryclass  = {hep-lat},
}

@Article{Chernodub:2011mc,
  author        = {Chernodub, M. N.},
  journal       = {Phys. Rev. Lett.},
  title         = {{Spontaneous electromagnetic superconductivity of vacuum in strong magnetic field: evidence from the Nambu--Jona-Lasinio model}},
  year          = {2011},
  pages         = {142003},
  volume        = {106},
  archiveprefix = {arXiv},
  doi           = {10.1103/PhysRevLett.106.142003},
  eprint        = {1101.0117},
  primaryclass  = {hep-ph},
}

@Article{Xu:2020yag,
  author        = {Xu, Kun and Chao, Jingyi and Huang, Mei},
  journal       = {Phys. Rev. D},
  title         = {{Effect of the anomalous magnetic moment of quarks on magnetized QCD matter and meson spectra}},
  year          = {2021},
  number        = {7},
  pages         = {076015},
  volume        = {103},
  archiveprefix = {arXiv},
  doi           = {10.1103/PhysRevD.103.076015},
  eprint        = {2007.13122},
  primaryclass  = {hep-ph},
}

@Article{Avancini:2015ady,
  author        = {Avancini, Sidney S. and Tavares, William R. and Pinto, Marcus B.},
  journal       = {Phys. Rev. D},
  title         = {{Properties of magnetized neutral mesons within a full RPA evaluation}},
  year          = {2016},
  number        = {1},
  pages         = {014010},
  volume        = {93},
  archiveprefix = {arXiv},
  doi           = {10.1103/PhysRevD.93.014010},
  eprint        = {1511.06261},
  primaryclass  = {hep-ph},
}

@Article{Brandt:2018bwq,
  author        = {Brandt, Bastian B. and Endrodi, Gergely and Fraga, Eduardo S. and Hippert, Mauricio and Schaffner-Bielich, Jurgen and Schmalzbauer, Sebastian},
  journal       = {Phys. Rev. D},
  title         = {{New class of compact stars: Pion stars}},
  year          = {2018},
  number        = {9},
  pages         = {094510},
  volume        = {98},
  archiveprefix = {arXiv},
  doi           = {10.1103/PhysRevD.98.094510},
  eprint        = {1802.06685},
  primaryclass  = {hep-ph},
}

@Article{Li:2020hlp,
  author        = {Li, Jianing and Cao, Gaoqing and He, Lianyi},
  journal       = {Phys. Rev. D},
  title         = {{Gauge independence of pion masses in a magnetic field within the Nambu{\textendash}Jona-Lasinio model}},
  year          = {2021},
  number        = {7},
  pages         = {074026},
  volume        = {104},
  archiveprefix = {arXiv},
  doi           = {10.1103/PhysRevD.104.074026},
  eprint        = {2009.04697},
  primaryclass  = {nucl-th},
}

@Article{Brauner:2016pko,
  author        = {Brauner, Tomas and Yamamoto, Naoki},
  journal       = {JHEP},
  title         = {{Chiral Soliton Lattice and Charged Pion Condensation in Strong Magnetic Fields}},
  year          = {2017},
  pages         = {132},
  volume        = {04},
  archiveprefix = {arXiv},
  doi           = {10.1007/JHEP04(2017)132},
  eprint        = {1609.05213},
  primaryclass  = {hep-ph},
}

@Article{Brauner:2016lkh,
  author        = {Brauner, Tomas and Huang, Xu-Guang},
  journal       = {Phys. Rev. D},
  title         = {{Vector meson condensation in a pion superfluid}},
  year          = {2016},
  number        = {9},
  pages         = {094003},
  volume        = {94},
  archiveprefix = {arXiv},
  doi           = {10.1103/PhysRevD.94.094003},
  eprint        = {1610.00426},
  primaryclass  = {hep-ph},
}

@Article{Colucci:2013zoa,
  author        = {Colucci, G. and Fraga, E. S. and Sedrakian, A.},
  journal       = {Phys. Lett. B},
  title         = {{Chiral pions in a magnetic background}},
  year          = {2014},
  pages         = {19--24},
  volume        = {728},
  archiveprefix = {arXiv},
  doi           = {10.1016/j.physletb.2013.11.028},
  eprint        = {1310.3742},
  primaryclass  = {nucl-th},
}

@Article{Luschevskaya:2014lga,
  author        = {Luschevskaya, E. V. and Solovjeva, O. E. and Kochetkov, O. A. and Teryaev, O. V.},
  journal       = {Nucl. Phys. B},
  title         = {{Magnetic polarizabilities of light mesons in $SU(3)$ lattice gauge theory}},
  year          = {2015},
  pages         = {627--643},
  volume        = {898},
  archiveprefix = {arXiv},
  doi           = {10.1016/j.nuclphysb.2015.07.023},
  eprint        = {1411.4284},
  primaryclass  = {hep-lat},
}

@Article{Klevansky:1992qe,
  author  = {Klevansky, S. P.},
  journal = {Rev. Mod. Phys.},
  title   = {{The Nambu-Jona-Lasinio model of quantum chromodynamics}},
  year    = {1992},
  pages   = {649--708},
  volume  = {64},
  doi     = {10.1103/RevModPhys.64.649},
}

@Unpublished{ConcurrentWork2026,
  author = {Liu, Zhiyang and Mao, Shijun},
  title  = {{Vector meson in magnetic field at finite temperature}},
  note   = {In preparation. While finalizing this work, we noted a concurrent investigation into the equivalence between the Schwinger proper time and Ritus formalisms. Our non-commutative Wigner-Weyl approach and their findings represent independent yet mutually reinforcing perspectives on the subject, having been developed without prior knowledge of each other.},
  year   = {2026}
}

@Article{Ritus:1972ky,
  author  = {Ritus, V. I.},
  journal = {Annals Phys.},
  title   = {{Radiative corrections in quantum electrodynamics with intense field and their analytical properties}},
  year    = {1972},
  pages   = {555--582},
  volume  = {69},
  doi     = {10.1016/0003-4916(72)90191-1},
}

@Article{Moyal:1949sk,
  author  = {Moyal, J. E.},
  journal = {Proc. Cambridge Phil. Soc.},
  title   = {{Quantum mechanics as a statistical theory}},
  year    = {1949},
  pages   = {99--124},
  volume  = {45},
  doi     = {10.1017/S0305004100000487},
}

@Article{Wen:2024hgu,
  author        = {Wen, Nanxiang and Cao, Xuanmin and Chao, Jingyi and Liu, Hui},
  journal       = {Phys. Rev. D},
  title         = {{Neutral pion masses within a hot and magnetized medium in a lattice-improved soft-wall AdS/QCD model}},
  year          = {2024},
  number        = {8},
  pages         = {086021},
  volume        = {109},
  archiveprefix = {arXiv},
  doi           = {10.1103/PhysRevD.109.086021},
  eprint        = {2402.06239},
  primaryclass  = {hep-th},
}

@Article{Yuan:2023dco,
  author        = {Yuan, Wen-Li and Chao, Jingyi and Li, Ang},
  journal       = {Phys. Rev. D},
  title         = {{Diquark and chiral condensates in a self-consistent NJL-type model}},
  year          = {2023},
  number        = {4},
  pages         = {043008},
  volume        = {108},
  archiveprefix = {arXiv},
  doi           = {10.1103/PhysRevD.108.043008},
  eprint        = {2304.12050},
  primaryclass  = {hep-ph},
}

@Article{Chao:2022bbv,
  author        = {Chao, Jingyi and Liu, Yu-Xin},
  journal       = {Phys. Rev. D},
  title         = {{Dimensional reduction and the generalized pion in a magnetic field within the NJL model}},
  year          = {2023},
  number        = {7},
  pages         = {074038},
  volume        = {107},
  archiveprefix = {arXiv},
  doi           = {10.1103/PhysRevD.107.074038},
  eprint        = {2202.05090},
  primaryclass  = {hep-ph},
}

@Article{Chao:2014wla,
  author        = {Chao, Jingyi and Yu, Lang and Huang, Mei},
  journal       = {Phys. Rev. D},
  title         = {{Zeta function regularization of the photon polarization tensor for a magnetized vacuum}},
  year          = {2014},
  note          = {[Erratum: Phys.Rev.D 91, 029903 (2015)]},
  number        = {4},
  pages         = {045033},
  volume        = {90},
  archiveprefix = {arXiv},
  doi           = {10.1103/PhysRevD.90.045033},
  eprint        = {1403.0442},
  primaryclass  = {hep-th},
}

@Article{Chao:2013qpa,
  author        = {Chao, Jingyi and Chu, Pengcheng and Huang, Mei},
  journal       = {Phys. Rev. D},
  title         = {{Inverse magnetic catalysis induced by sphalerons}},
  year          = {2013},
  pages         = {054009},
  volume        = {88},
  archiveprefix = {arXiv},
  doi           = {10.1103/PhysRevD.88.054009},
  eprint        = {1305.1100},
  primaryclass  = {hep-ph},
}

@Article{Liu:2014uwa,
  author        = {Liu, Hao and Yu, Lang and Huang, Mei},
  journal       = {Phys. Rev. D},
  title         = {{Charged and neutral vector $\rho$ mesons in a magnetic field}},
  year          = {2015},
  number        = {1},
  pages         = {014017},
  volume        = {91},
  archiveprefix = {arXiv},
  doi           = {10.1103/PhysRevD.91.014017},
  eprint        = {1408.1318},
  primaryclass  = {hep-ph},
}

@Article{Berra-Montiel:2024ubb,
  author        = {Berra-Montiel, Jasel and Garcia-Compean, Hugo and Molgado, Alberto},
  journal       = {Annals Phys.},
  title         = {{Star exponentials from propagators and path integrals}},
  year          = {2024},
  pages         = {169744},
  volume        = {468},
  archiveprefix = {arXiv},
  doi           = {10.1016/j.aop.2024.169744},
  eprint        = {2404.08815},
  primaryclass  = {math-ph},
}

@Book{Gradshteyn:1943cpj,
  author = {Gradshteyn, I. S. and Ryzhik, I. M.},
  title  = {{Table of Integrals, Series, and Products}},
  year   = {1943},
  isbn   = {978-0-12-294757-5, 978-0-12-294757-5},
}

@Article{Szabo:2001kg,
  author        = {Szabo, Richard J.},
  journal       = {Phys. Rept.},
  title         = {{Quantum field theory on noncommutative spaces}},
  year          = {2003},
  pages         = {207--299},
  volume        = {378},
  archiveprefix = {arXiv},
  doi           = {10.1016/S0370-1573(03)00059-0},
  eprint        = {hep-th/0109162},
  reportnumber  = {HWM-01-35, EMPG-01-14},
}

@Article{Langmann:2003cg,
  author        = {Langmann, E. and Szabo, R. J. and Zarembo, K.},
  journal       = {Phys. Lett. B},
  title         = {{Exact solution of noncommutative field theory in background magnetic fields}},
  year          = {2003},
  pages         = {95--101},
  volume        = {569},
  archiveprefix = {arXiv},
  doi           = {10.1016/j.physletb.2003.07.020},
  eprint        = {hep-th/0303082},
  reportnumber  = {HWM-03-6, EMPG-03-06, UUITP-02-03},
}

@Article{DElia:2011koc,
  author        = {D'Elia, Massimo and Negro, Francesco},
  journal       = {Phys. Rev. D},
  title         = {{Chiral Properties of Strong Interactions in a Magnetic Background}},
  year          = {2011},
  pages         = {114028},
  volume        = {83},
  archiveprefix = {arXiv},
  doi           = {10.1103/PhysRevD.83.114028},
  eprint        = {1103.2080},
  primaryclass  = {hep-lat},
}

@Article{Ding:2020inp,
  author        = {Ding, Heng-Tong and Schmidt, Christian and Tomiya, Akio and Wang, Xiao-Dan},
  journal       = {Phys. Rev. D},
  title         = {{Chiral phase structure of three flavor QCD in a background magnetic field}},
  year          = {2020},
  number        = {5},
  pages         = {054505},
  volume        = {102},
  archiveprefix = {arXiv},
  doi           = {10.1103/PhysRevD.102.054505},
  eprint        = {2006.13422},
  primaryclass  = {hep-lat},
}

@Article{Endrodi:2024cqn,
  author        = {Endrodi, Gergely},
  journal       = {Prog. Part. Nucl. Phys.},
  title         = {{QCD with background electromagnetic fields on the lattice: A review}},
  year          = {2025},
  pages         = {104153},
  volume        = {141},
  archiveprefix = {arXiv},
  doi           = {10.1016/j.ppnp.2024.104153},
  eprint        = {2406.19780},
  primaryclass  = {hep-lat},
}

@Article{Orlovsky:2013gha,
  author        = {Orlovsky, V. D. and Simonov, Yu. A.},
  journal       = {JHEP},
  title         = {{Nambu-Goldstone mesons in strong magnetic field}},
  year          = {2013},
  pages         = {136},
  volume        = {09},
  archiveprefix = {arXiv},
  doi           = {10.1007/JHEP09(2013)136},
  eprint        = {1306.2232},
  primaryclass  = {hep-ph},
}

@Article{Shovkovy:2012zn,
  author        = {Shovkovy, Igor A.},
  journal       = {Lect. Notes Phys.},
  title         = {{Magnetic Catalysis: A Review}},
  year          = {2013},
  pages         = {13--49},
  volume        = {871},
  archiveprefix = {arXiv},
  doi           = {10.1007/978-3-642-37305-3_2},
  eprint        = {1207.5081},
  primaryclass  = {hep-ph},
}

@Article{Kharzeev:2012ph,
  author        = {Kharzeev, Dmitri E. and Landsteiner, Karl and Schmitt, Andreas and Yee, Ho-Ung},
  journal       = {Lect. Notes Phys.},
  title         = {{'Strongly interacting matter in magnetic fields': an overview}},
  year          = {2013},
  pages         = {1--11},
  volume        = {871},
  archiveprefix = {arXiv},
  doi           = {10.1007/978-3-642-37305-3_1},
  eprint        = {1211.6245},
  primaryclass  = {hep-ph},
}

@Article{Bali:2018sey,
  author        = {Bali, G. S. and Brandt, B. B. and Endr\H{o}di, G. and Gl\"a\ss{}le, B.},
  journal       = {Phys. Rev. Lett.},
  title         = {{Weak decay of magnetized pions}},
  year          = {2018},
  number        = {7},
  pages         = {072001},
  volume        = {121},
  archiveprefix = {arXiv},
  doi           = {10.1103/PhysRevLett.121.072001},
  eprint        = {1805.10971},
  primaryclass  = {hep-lat},
}

@Article{Avancini:2016fgq,
  author        = {Avancini, Sidney S. and Farias, Ricardo L. S. and Benghi Pinto, Marcus and Tavares, William R. and Tim\'oteo, Varese S.},
  journal       = {Phys. Lett. B},
  title         = {{$\pi_0$ pole mass calculation in a strong magnetic field and lattice constraints}},
  year          = {2017},
  pages         = {247--252},
  volume        = {767},
  archiveprefix = {arXiv},
  doi           = {10.1016/j.physletb.2017.02.002},
  eprint        = {1606.05754},
  primaryclass  = {hep-ph},
}

@Article{Avancini:2018svs,
  author        = {Avancini, Sidney S. and Farias, Ricardo L. S. and Tavares, William R.},
  journal       = {Phys. Rev. D},
  title         = {{Neutral meson properties in hot and magnetized quark matter: a new magnetic field independent regularization scheme applied to NJL-type model}},
  year          = {2019},
  number        = {5},
  pages         = {056009},
  volume        = {99},
  archiveprefix = {arXiv},
  doi           = {10.1103/PhysRevD.99.056009},
  eprint        = {1812.00945},
  primaryclass  = {hep-ph},
}

@Article{Chernodub:2010qx,
  author        = {M. N. Chernodub},
  journal       = {Phys. Rev. D},
  title         = {{Superconductivity of QCD vacuum in strong magnetic field}},
  year          = {2010},
  pages         = {085011},
  volume        = {82},
  archiveprefix = {arXiv},
  doi           = {10.1103/PhysRevD.82.085011},
  eprint        = {1008.1055},
  primaryclass  = {hep-ph},
}

@Misc{Kojo:2026gep,
  author        = {Kojo, Toru and Itatani, Sakura},
  title         = {{Delineating neutral and charged mesons in magnetic fields}},
  year          = {2026},
  month         = {4},
  archiveprefix = {arXiv},
  eprint        = {2604.15897},
  primaryclass  = {hep-ph},
  reportnumber  = {KEK-TH-2822},
}

@Article{Kharzeev:2015znc,
  author        = {Kharzeev, D. E. and Liao, J. and Voloshin, S. A. and Wang, G.},
  journal       = {Prog. Part. Nucl. Phys.},
  title         = {{Chiral magnetic and vortical effects in high-energy nuclear collisions{\textemdash}A status report}},
  year          = {2016},
  pages         = {1--28},
  volume        = {88},
  archiveprefix = {arXiv},
  doi           = {10.1016/j.ppnp.2016.01.001},
  eprint        = {1511.04050},
  primaryclass  = {hep-ph},
}

@Article{Fukushima:2012kc,
  author        = {Fukushima, Kenji and Hidaka, Yoshimasa},
  journal       = {Phys. Rev. Lett.},
  title         = {{Magnetic Catalysis Versus Magnetic Inhibition}},
  year          = {2013},
  number        = {3},
  pages         = {031601},
  volume        = {110},
  archiveprefix = {arXiv},
  doi           = {10.1103/PhysRevLett.110.031601},
  eprint        = {1209.1319},
  primaryclass  = {hep-ph},
}

@Article{Cao:2021rwx,
  author        = {Cao, Gaoqing},
  journal       = {Eur. Phys. J. A},
  title         = {{Recent progresses on QCD phases in a strong magnetic field: views from Nambu{\textendash}Jona-Lasinio model}},
  year          = {2021},
  number        = {9},
  pages         = {264},
  volume        = {57},
  archiveprefix = {arXiv},
  doi           = {10.1140/epja/s10050-021-00570-0},
  eprint        = {2103.00456},
  primaryclass  = {hep-ph},
}

@Misc{Wang:2026xsm,
  author        = {Wang, Ziyue},
  title         = {{Residue-Enhanced Pion-Rho Mixing as the Origin of Nonmonotonic Charged Pion Mass in Magnetic Fields}},
  year          = {2026},
  month         = {2},
  archiveprefix = {arXiv},
  eprint        = {2602.15410},
  primaryclass  = {hep-ph},
}

@Article{Andreichikov:2016ayj,
  author        = {Andreichikov, M. A. and Kerbikov, B. O. and Luschevskaya, E. V. and Simonov, Yu. A. and Solovjeva, O. E.},
  journal       = {JHEP},
  title         = {{The Evolution of Meson Masses in a Strong Magnetic Field}},
  year          = {2017},
  pages         = {007},
  volume        = {05},
  archiveprefix = {arXiv},
  doi           = {10.1007/JHEP05(2017)007},
  eprint        = {1610.06887},
  primaryclass  = {hep-ph},
}

@Article{Coppola:2023mmq,
  author        = {Coppola, M{\'a}ximo and Gomez Dumm, Daniel and Noguera, Santiago and Scoccola, Norberto N.},
  journal       = {Phys. Rev. D},
  title         = {{Masses of magnetized pseudoscalar and vector mesons in an extended NJL model: The role of axial vector mesons}},
  year          = {2024},
  number        = {5},
  pages         = {054014},
  volume        = {109},
  archiveprefix = {arXiv},
  doi           = {10.1103/PhysRevD.109.054014},
  eprint        = {2312.16675},
  primaryclass  = {hep-ph},
}

@Article{Carlomagno:2022arc,
  author        = {Carlomagno, J. P. and Gomez Dumm, D. and Villafa{\~n}e, M. F. Izzo and Noguera, S. and Scoccola, N. N.},
  journal       = {Phys. Rev. D},
  title         = {{Charged pseudoscalar and vector meson masses in strong magnetic fields in an extended NJL model}},
  year          = {2022},
  number        = {9},
  pages         = {094035},
  volume        = {106},
  archiveprefix = {arXiv},
  doi           = {10.1103/PhysRevD.106.094035},
  eprint        = {2209.10679},
  primaryclass  = {hep-ph},
}

@Article{Skokov:2009qp,
  author        = {Skokov, V. and Illarionov, A. Yu. and Toneev, V.},
  journal       = {Int. J. Mod. Phys. A},
  title         = {{Estimate of the magnetic field strength in heavy-ion collisions}},
  year          = {2009},
  pages         = {5925--5932},
  volume        = {24},
  archiveprefix = {arXiv},
  doi           = {10.1142/S0217751X09047570},
  eprint        = {0907.1396},
  primaryclass  = {nucl-th},
}

@Article{Deng:2012pc,
  author        = {Deng, Wei-Tian and Huang, Xu-Guang},
  journal       = {Phys. Rev. C},
  title         = {{Event-by-event generation of electromagnetic fields in heavy-ion collisions}},
  year          = {2012},
  pages         = {044907},
  volume        = {85},
  archiveprefix = {arXiv},
  doi           = {10.1103/PhysRevC.85.044907},
  eprint        = {1201.5108},
  primaryclass  = {nucl-th},
}

@Article{Mao:2018dqe,
  author        = {Mao, Shijun},
  journal       = {Phys. Rev. D},
  title         = {{Pions in magnetic field at finite temperature}},
  year          = {2019},
  number        = {5},
  pages         = {056005},
  volume        = {99},
  archiveprefix = {arXiv},
  doi           = {10.1103/PhysRevD.99.056005},
  eprint        = {1808.10242},
  primaryclass  = {nucl-th},
}

@Article{Coppola:2019uyr,
  author        = {Coppola, M. and Gomez Dumm, D. and Noguera, S. and Scoccola, N. N.},
  journal       = {Phys. Rev. D},
  title         = {{Neutral and charged pion properties under strong magnetic fields in the NJL model}},
  year          = {2019},
  number        = {5},
  pages         = {054014},
  volume        = {100},
  archiveprefix = {arXiv},
  doi           = {10.1103/PhysRevD.100.054014},
  eprint        = {1907.05840},
  primaryclass  = {hep-ph},
}

@Article{Miransky:2015ava,
  author        = {Miransky, Vladimir A. and Shovkovy, Igor A.},
  journal       = {Phys. Rept.},
  title         = {{Quantum field theory in a magnetic field: From quantum chromodynamics to graphene and Dirac semimetals}},
  year          = {2015},
  pages         = {1--209},
  volume        = {576},
  archiveprefix = {arXiv},
  doi           = {10.1016/j.physrep.2015.02.003},
  eprint        = {1503.00732},
  primaryclass  = {hep-ph},
}

@Article{Gusynin:1995nb,
  author        = {Gusynin, V. P. and Miransky, V. A. and Shovkovy, I. A.},
  journal       = {Nucl. Phys. B},
  title         = {{Dimensional reduction and catalysis of dynamical symmetry breaking by a magnetic field}},
  year          = {1996},
  pages         = {249--290},
  volume        = {462},
  archiveprefix = {arXiv},
  doi           = {10.1016/0550-3213(96)00021-1},
  eprint        = {hep-ph/9509320},
  reportnumber  = {UCLA-95-TEP-26},
}

@Article{Hidaka:2012mz,
  author        = {Hidaka, Yoshimasa and Yamamoto, Arata},
  journal       = {Phys. Rev. D},
  title         = {{Charged vector mesons in a strong magnetic field}},
  year          = {2013},
  number        = {9},
  pages         = {094502},
  volume        = {87},
  archiveprefix = {arXiv},
  doi           = {10.1103/PhysRevD.87.094502},
  eprint        = {1209.0007},
  primaryclass  = {hep-ph},
  reportnumber  = {RIKEN-QHP-44},
}

@Article{Bali:2012zg,
  author        = {Bali, G. S. and Bruckmann, F. and Endrodi, G. and Fodor, Z. and Katz, S. D. and Schafer, A.},
  journal       = {Phys. Rev. D},
  title         = {{QCD quark condensate in external magnetic fields}},
  year          = {2012},
  pages         = {071502},
  volume        = {86},
  archiveprefix = {arXiv},
  doi           = {10.1103/PhysRevD.86.071502},
  eprint        = {1206.4205},
  primaryclass  = {hep-lat},
}

@Article{Bruckmann:2013oba,
  author        = {Bruckmann, Falk and Endrodi, Gergely and Kovacs, Tamas G.},
  journal       = {JHEP},
  title         = {{Inverse magnetic catalysis and the Polyakov loop}},
  year          = {2013},
  pages         = {112},
  volume        = {04},
  archiveprefix = {arXiv},
  doi           = {10.1007/JHEP04(2013)112},
  eprint        = {1303.3972},
  primaryclass  = {hep-lat},
}

@Article{Kojo:2021gvm,
  author        = {Kojo, Toru},
  journal       = {Eur. Phys. J. A},
  title         = {{Neutral and charged mesons in magnetic fields: A resonance gas in a non-relativistic quark model}},
  year          = {2021},
  number        = {11},
  pages         = {317},
  volume        = {57},
  archiveprefix = {arXiv},
  doi           = {10.1140/epja/s10050-021-00629-y},
  eprint        = {2104.00376},
  primaryclass  = {hep-ph},
}

@Article{Fukushima:2008xe,
  author        = {Fukushima, Kenji and Kharzeev, Dmitri E. and Warringa, Harmen J.},
  journal       = {Phys. Rev. D},
  title         = {{The Chiral Magnetic Effect}},
  year          = {2008},
  pages         = {074033},
  volume        = {78},
  archiveprefix = {arXiv},
  doi           = {10.1103/PhysRevD.78.074033},
  eprint        = {0808.3382},
  primaryclass  = {hep-ph},
}

@Article{Bali:2017ian,
  author        = {Bali, Gunnar S. and Brandt, Bastian B. and Endr\H{o}di, Gergely and Gl\"a\ss{}le, Benjamin},
  journal       = {Phys. Rev. D},
  title         = {{Meson masses in electromagnetic fields with Wilson fermions}},
  year          = {2018},
  number        = {3},
  pages         = {034505},
  volume        = {97},
  archiveprefix = {arXiv},
  doi           = {10.1103/PhysRevD.97.034505},
  eprint        = {1707.05600},
  primaryclass  = {hep-lat},
}

@Article{Andersen:2014xxa,
  author        = {Andersen, Jens O. and Naylor, William R. and Tranberg, Anders},
  journal       = {Rev. Mod. Phys.},
  title         = {{Phase diagram of QCD in a magnetic field: A review}},
  year          = {2016},
  pages         = {025001},
  volume        = {88},
  archiveprefix = {arXiv},
  doi           = {10.1103/RevModPhys.88.025001},
  eprint        = {1411.7176},
  primaryclass  = {hep-ph},
}

@Article{Li:2016gfn,
  author        = {Li, Danning and Huang, Mei and Yang, Yi and Yuan, Pei-Hung},
  journal       = {JHEP},
  title         = {{Inverse Magnetic Catalysis in the Soft-Wall Model of AdS/QCD}},
  year          = {2017},
  pages         = {030},
  volume        = {02},
  archiveprefix = {arXiv},
  doi           = {10.1007/JHEP02(2017)030},
  eprint        = {1610.04618},
  primaryclass  = {hep-th},
}

@Article{Avancini:2021pmi,
  author        = {Avancini, Sidney S. and Coppola, M{\'a}ximo and Scoccola, Norberto N. and Sodr{\'e}, Joana C.},
  journal       = {Phys. Rev. D},
  title         = {{Light pseudoscalar meson masses under strong magnetic fields within the SU(3) Nambu{\textendash}Jona-Lasinio model}},
  year          = {2021},
  number        = {9},
  pages         = {094040},
  volume        = {104},
  archiveprefix = {arXiv},
  doi           = {10.1103/PhysRevD.104.094040},
  eprint        = {2109.01911},
  primaryclass  = {hep-ph},
}

@Article{Wen:2023qcz,
  author        = {Wen, Rui and Yin, Shi and Fu, Wei-jie and Huang, Mei},
  journal       = {Phys. Rev. D},
  title         = {{Functional renormalization group study of neutral and charged pions in magnetic fields in the quark-meson model}},
  year          = {2023},
  number        = {7},
  pages         = {076020},
  volume        = {108},
  archiveprefix = {arXiv},
  doi           = {10.1103/PhysRevD.108.076020},
  eprint        = {2306.04045},
  primaryclass  = {hep-ph},
}

@Article{Ghosh:2016evc,
  author        = {Ghosh, Snigdha and Mukherjee, Arghya and Mandal, Mahatsab and Sarkar, Sourav and Roy, Pradip},
  journal       = {Phys. Rev. D},
  title         = {{Spectral properties of $\rho$ meson in a magnetic field}},
  year          = {2016},
  number        = {9},
  pages         = {094043},
  volume        = {94},
  archiveprefix = {arXiv},
  doi           = {10.1103/PhysRevD.94.094043},
  eprint        = {1612.02966},
  primaryclass  = {nucl-th},
}

@Article{Kawaguchi:2015gpt,
  author        = {Kawaguchi, Mamiya and Matsuzaki, Shinya},
  journal       = {Phys. Rev. D},
  title         = {{Vector meson masses from a hidden local symmetry in a constant magnetic field}},
  year          = {2016},
  number        = {12},
  pages         = {125027},
  volume        = {93},
  archiveprefix = {arXiv},
  doi           = {10.1103/PhysRevD.93.125027},
  eprint        = {1511.06990},
  primaryclass  = {hep-ph},
}

@Article{GomezDumm:2023owj,
  author        = {Gomez Dumm, D. and Noguera, S. and Scoccola, N. N.},
  journal       = {Phys. Rev. D},
  title         = {Charged meson masses under strong magnetic fields: Gauge invariance and Schwinger phases},
  year          = {2023},
  number        = {1},
  pages         = {016012},
  volume        = {108},
  archiveprefix = {arXiv},
  doi           = {10.1103/PhysRevD.108.016012},
  eprint        = {2306.04128},
  primaryclass  = {hep-ph},
}

@Article{Ding:2025zqu,
  author        = {Ding, Minghui and Gao, Fei and Schmidt, Sebastian M.},
  journal       = {Chin. Phys.},
  title         = {{Anisotropic quark propagation and Zeeman effect in an external magnetic field*}},
  year          = {2026},
  number        = {1},
  pages         = {013104},
  volume        = {50},
  archiveprefix = {arXiv},
  doi           = {10.1088/1674-1137/ae07b9},
  eprint        = {2504.14504},
  primaryclass  = {hep-ph},
}

@Article{Sheng:2022ssp,
  author        = {Sheng, Xin-Li and Yang, Shu-Yun and Zou, Yao-Lin and Hou, Defu},
  journal       = {Eur. Phys. J. C},
  title         = {{Mass splitting and spin alignment for $\phi $ mesons in a magnetic field in NJL model}},
  year          = {2024},
  number        = {3},
  pages         = {299},
  volume        = {84},
  archiveprefix = {arXiv},
  doi           = {10.1140/epjc/s10052-024-12643-7},
  eprint        = {2209.01872},
  primaryclass  = {nucl-th},
}

@Article{Ding:2020hxw,
  author        = {Ding, H. -T. and Li, S. -T. and Tomiya, A. and Wang, X. -D. and Zhang, Y.},
  journal       = {Phys. Rev. D},
  title         = {{Chiral properties of (2+1)-flavor QCD in strong magnetic fields at zero temperature}},
  year          = {2021},
  number        = {1},
  pages         = {014505},
  volume        = {104},
  archiveprefix = {arXiv},
  doi           = {10.1103/PhysRevD.104.014505},
  eprint        = {2008.00493},
  primaryclass  = {hep-lat},
}

@Article{Li:2023rsy,
  author        = {Li, Luyang and Mao, Shijun},
  journal       = {Phys. Rev. D},
  title         = {{Inverse magnetic catalysis effect and current quark mass effect on mass spectra and Mott transitions of pions under external magnetic field}},
  year          = {2023},
  number        = {5},
  pages         = {054001},
  volume        = {108},
  archiveprefix = {arXiv},
  doi           = {10.1103/PhysRevD.108.054001},
  eprint        = {2308.12491},
  primaryclass  = {hep-ph},
}

@Misc{Ding:2026qzu,
  author        = {Ding, Heng-Tong and Zhang, Dan},
  title         = {{Chiral Properties of $(2\!+\!1)$-Flavor QCD in Magnetic Fields at Zero Temperature}},
  year          = {2026},
  month         = {1},
  archiveprefix = {arXiv},
  eprint        = {2601.18354},
  primaryclass  = {hep-lat},
}

@Article{Zachos:1999wn,
  author        = {Zachos, Cosmas K. and Curtright, Thomas},
  journal       = {Prog. Theor. Phys. Suppl.},
  title         = {{Phase space quantization of field theory}},
  year          = {1999},
  pages         = {244--258},
  volume        = {135},
  archiveprefix = {arXiv},
  doi           = {10.1143/PTPS.135.244},
  editor        = {Inami, T. and Sasaki, R. and Uematsu, T.},
  eprint        = {hep-th/9903254},
  reportnumber  = {ANL-HEP-CP-99-06, MIAMI-TH-1-99, ANL-HEP-CP-99-06, Miami TH/1/99},
}

@Article{Curtright:1997me,
  author        = {Curtright, Thomas and Fairlie, David and Zachos, Cosmas K.},
  journal       = {Phys. Rev. D},
  title         = {{Features of time independent Wigner functions}},
  year          = {1998},
  pages         = {025002},
  volume        = {58},
  archiveprefix = {arXiv},
  doi           = {10.1103/PhysRevD.58.025002},
  eprint        = {hep-th/9711183},
  reportnumber  = {DTP-97-61, MIAMI-TH-97-3, ANL-HEP-PR-97-93, DTP/97/61, MIAMI-TH-97-3, ANL-HEP-PR-97-93},
}

@Article{Coppola:2018vkw,
  author        = {Coppola, M. and G{\'o}mez Dumm, D. and Scoccola, N. N.},
  journal       = {Phys. Lett. B},
  title         = {{Charged pion masses under strong magnetic fields in the NJL model}},
  year          = {2018},
  pages         = {155--161},
  volume        = {782},
  archiveprefix = {arXiv},
  doi           = {10.1016/j.physletb.2018.04.043},
  eprint        = {1802.08041},
  primaryclass  = {hep-ph},
}

@Article{Callebaut:2010mct,
  author        = {Callebaut, Nele and Dudal, David and Verschelde, Henri},
  journal       = {PoS},
  title         = {{Holographic study of rho meson mass in an external magnetic field: Paving the road towards a magnetically induced superconducting QCD vacuum?}},
  year          = {2010},
  pages         = {046},
  volume        = {FACESQCD},
  archiveprefix = {arXiv},
  doi           = {10.22323/1.117.0046},
  eprint        = {1102.3103},
  primaryclass  = {hep-ph},
}

@Article{Xing:2021kbw,
  author        = {Xing, Zanbin and Chao, Jingyi and Chang, Lei and Liu, Yu-xin},
  journal       = {Phys. Rev. D},
  title         = {{Exposing the effect of the p-wave component in the pion triplet under a strong magnetic field}},
  year          = {2022},
  number        = {11},
  pages         = {114003},
  volume        = {105},
  archiveprefix = {arXiv},
  doi           = {10.1103/PhysRevD.105.114003},
  eprint        = {2110.01245},
  primaryclass  = {hep-ph},
}

@article{Braby:2009dw,
    author = {Braby, Matt and Chao, Jingyi and Sch{\"a}fer, Thomas},
    title = "{Thermal conductivity of color-flavor locked quark matter}",
    eprint = "0909.4236",
    archivePrefix = "arXiv",
    primaryClass = "hep-ph",
    doi = "10.1103/PhysRevC.81.045205",
    journal = "Phys. Rev. C",
    volume = "81",
    pages = "045205",
    year = "2010"
}

@article{Braby:2010tk,
    author = {Braby, Matt and Chao, Jingyi and Sch{\"a}fer, Thomas},
    title = "{Viscosity spectral functions of the dilute Fermi gas in kinetic theory}",
    eprint = "1012.0219",
    archivePrefix = "arXiv",
    primaryClass = "cond-mat.quant-gas",
    doi = "10.1088/1367-2630/13/3/035014",
    journal = "New J. Phys.",
    volume = "13",
    pages = "035014",
    year = "2011"
}

@article{Huang:2015oca,
    author = "Huang, Xu-Guang",
    title = "{Electromagnetic fields and anomalous transports in heavy-ion collisions --- A pedagogical review}",
    eprint = "1509.04073",
    archivePrefix = "arXiv",
    primaryClass = "nucl-th",
    doi = "10.1088/0034-4885/79/7/076302",
    journal = "Rept. Prog. Phys.",
    volume = "79",
    number = "7",
    pages = "076302",
    year = "2016"
}

@Misc{Mei:2026xlj,
    author = "Mei, Jie and Wen, Rui and Zhou, Min and Mao, Shijun and Huang, Mei",
    title = "{Spectral function for pions in magnetic field}",
    eprint = "2601.22422",
    archivePrefix = "arXiv",
    primaryClass = "hep-ph",
    month = "1",
    year = "2026"
}

@article{Ghosh:2020qvg,
    author = "Ghosh, Snigdha and Mukherjee, Arghya and Chaudhuri, Nilanjan and Roy, Pradip and Sarkar, Sourav",
    title = "{Thermo-magnetic spectral properties of neutral mesons in vector and axial-vector channels using NJL model}",
    eprint = "2003.02024",
    archivePrefix = "arXiv",
    primaryClass = "hep-ph",
    doi = "10.1103/PhysRevD.101.056023",
    journal = "Phys. Rev. D",
    volume = "101",
    number = "5",
    pages = "056023",
    year = "2020"
}

@article{Chang:2009zb,
    author = "Chang, Lei and Roberts, Craig D.",
    title = "{Sketching the Bethe-Salpeter kernel}",
    eprint = "0903.5461",
    archivePrefix = "arXiv",
    primaryClass = "nucl-th",
    doi = "10.1103/PhysRevLett.103.081601",
    journal = "Phys. Rev. Lett.",
    volume = "103",
    pages = "081601",
    year = "2009"
}

@article{Sheng:2017lfu,
    author = "Sheng, Xin-li and Rischke, Dirk H. and Vasak, David and Wang, Qun",
    title = "{Wigner functions for fermions in strong magnetic fields}",
    eprint = "1707.01388",
    archivePrefix = "arXiv",
    primaryClass = "hep-ph",
    reportNumber = "ICTS-USTC-17-12",
    doi = "10.1140/epja/i2018-12414-9",
    journal = "Eur. Phys. J. A",
    volume = "54",
    number = "2",
    pages = "21",
    year = "2018"
}

@article{Fang:2023bbw,
    author = "Fang, Shuo and Pu, Shi and Yang, Di-Lun",
    title = "{Spin polarization and spin alignment from quantum kinetic theory with self-energy corrections}",
    eprint = "2311.15197",
    archivePrefix = "arXiv",
    primaryClass = "hep-ph",
    doi = "10.1103/PhysRevD.109.034034",
    journal = "Phys. Rev. D",
    volume = "109",
    number = "3",
    pages = "034034",
    year = "2024"
}
\end{document}